\documentclass[aps,prc,twocolumn,floatfix,superscriptaddress]{revtex4}
\usepackage{epsfig}
\usepackage{amsmath}
\usepackage{color}
\usepackage[utf8]{inputenc}
\usepackage{graphicx}

\definecolor{blue}{rgb}{0.05, 0.05, 0.5}

\begin{document}

\title{Transport Dynamics of Parton Interactions in pp Collisions  at LHC Energies}
\author{Dinesh K. Srivastava}
\email{dinesh@vecc.gov.in}
\affiliation{Variable Energy Cyclotron Centre, HBNI, 1/AF, Bidhan Nagar, Kolkata 700064, India}
\affiliation{Institut für Theoretische Physik, Johann Wolfgang Goethe-Universität, Max-von-Laue-Str. 1, D-60438 Frankfurt am Main, Germany}
\affiliation{ExtreMe Matter Institute EMMI, GSI Helmholtzzentrum für Schwerionenforschung, Planckstrasse 1, 64291 Darmstadt, Germany}
\author{Rupa Chatterjee}
\email{rupa@vecc.gov.in}
\affiliation{Variable Energy Cyclotron Centre, HBNI, 1/AF, Bidhan Nagar, Kolkata 700064, India}
\author{Steffen A. Bass}
\email{bass@phy.duke.edu}
\affiliation{Duke University, Dept. of Physics, 139 Science Drive,
Box 90305, Durham NC 27708, U. S. A.}

\begin{abstract}
We investigate the transport dynamics of partons in proton-proton collisions at the Large Hadron Collider using a Boltzmann transport approach, the parton cascade model.  The calculations include semi-hard pQCD interaction of partons populating 
the nucleons and provide a space-time description of the collision in terms of cascading partons
undergoing scatterings and fragmentations.
Parton production and number of collisions rise rapidly with increase in center of mass energy of the collision.
For a given center of mass energy, the number of parton interactions is seen to
rise stronger than linear with decreasing impact parameter before saturating for very central collisions. The strangeness enhancement factor $\gamma_s$ 
for the semi-hard processes is found to rise rapidly and
saturate towards the highest collision energies. Overall, our study indicates a significant amount of partonic interactions in proton-proton collisions, which supports the observation of fluid-like behavior for high multiplicity proton-proton collisions observed in the experiments.

\end{abstract}


\maketitle

\maketitle
\section{Introduction} 
Relativistic heavy-ion collisions have been used with great success to probe the properties of hot and dense QCD matter, the quark-gluon-plasma (QGP) \cite{Arsene:2004fa,Adcox:2004mh,Back:2004je,Adams:2005dq,Gyulassy:2004zy,Muller:2006ee,Muller:2012zq}. Proton-proton collisions at a given center of mass energy per nucleon have been thought to provide a baseline measurement without the creation of a QGP when extrapolated to the corresponding nucleus-nucleus case using simple geometric models \cite{Miller:2007ri}. However, recently this canonical picture has undergone a considerable change
as several experimental "indications" of formation of a medium, e.g., flow and enhanced production of strangeness have also been seen in proton-proton collisions, albeit only when triggering on high multiplicity events \cite{ALICE:2017jyt}. 

One should note that on the theory side the notion of possible QGP formation in proton-proton collisions dates back several decades:
hydrodynamics has been used for a long time while exploring $pp$ collisions, with several theoretical studies
assuming formation of QGP (see, e.g., Ref.~\cite{Shuryak:1978ij,VonGersdorff:1986tqh,McLerran:1986nc}).
It was suggested that the particle spectra for 1.8 TeV proton-antiproton collisions showed
evidence of flow~\cite{Levai:1991be} which indicated formation of quark-gluon plasma in such collisions.
One may recall though, that it was argued~\cite{Wang:1991vx} that the increase in $\langle p_T^2 \rangle$ 
could be attributed to events simply having a large multiplicity, i.e. increased minijet activity, without the formation of a deconfined medium.
Subsequently, additional data for the same system, along with measurements of HBT radii, were used~\cite{Alexopoulos:2002eh} to claim evidence for a deconfining phase transition in these collisions. However, only through the advent of recent high quality data have these calculations and analyses gained renewed traction. By now a large body of data for $pp$ collisions at RHIC and ever-increasing LHC energies has been accumulated,
which provides enough indications for flow and enhanced production of strangeness in events having large 
multiplicity~\cite{ALICE:2017jyt}, even though non-QGP based interpretations of the data have remained viable \cite{Blok:2017pui,Greif:2017dua,Aichelin:2016vpr}.

In the present work we aim at quantifying the amount of parton interactions and rescattering present in proton-proton collisions and whether the amount of interactions observed may lend credence to the notion of collectivity in these collision-systems and the application of hydrodynamic models. For this purpose we use a microscopic Boltzmann transport approach, the parton cascade model~\cite{Geiger:1991nj} as implemented in VNI/BMS~\cite{Bass:2002fh}
and extended recently to include heavy quark production~\cite{Younus:2013rja,Srivastava:2017bcm} to explore the emergence of semi-hard multi-partonic
collisions and parton multiplications in  $pp$ collisions using pQCD matrix elements.
This treatment~\cite{Bass:2002fh} has several inherent advantages.
First of all, all parton scatterings leading to $p_T \ge p_T^\text{cut-off}$ are treated within (lowest order)
perturbative QCD, avoiding any arbitrariness, except for the dependence of the results on the momentum cut-off, introduced
to avoid singular cross-sections for mass-less partons at lower momentum transfers. However, it is
expected that spectra etc. for larger $p_T$ should be reasonable. It should be noted, however, that the limitation to pQCD matrix elements with a momentum cut-off implies that our approach does not describe the dynamics of thermalized degrees of freedom. We thus will only be able to assess whether the conditions necessary for the formation of (equilibrated) QCD matter are met, but will not be able to describe the development and evolution of a QGP itself.

The tracking of the hard collision dynamics and all the partons involved in these interactions  allows us to perform calculations at several levels of complexity: in a first step, we only allow the primary partons from the projectile nucleon to
collide with  primary partons from the target nucleon. Next we consider scattering among 
primary and secondary partons. This corresponds~\cite{Bass:2002fh} most closely to minijet 
calculations~\cite{Eichten:1984eu}.
Finally we perform calculations which account for fragmentation of final state partons following semi-hard scatterings.
These radiative processes are included following the
original PCM implementation~\cite{Geiger:1991nj} in the leading-logarithmic approximation (LLA).

We do not consider hadronization of either the partons which have undergone interaction
 or the un-interacted partons, and thus our findings
relate only to the partons produced in the semi-hard processes.

We report our results for minimum bias collisions of protons at center of mass energies of
 0.2, 2.76, 5.02, 7.00, and 14 TeV for the three implementations discussed above and study the evolution of
parton production and multiple collisions with the increase in collision energy. Subsequently
we explore the collision of protons at a center of mass energy of 7.00 TeV as a function of impact parameter and the $p_T^{\text {cut-off}}$ used for regularizing the infra-red divergences for pQCD cross-sections. Finally, we study some of these systematics as a function of number of quarks (charged particles).

\begin{figure}
\centerline{\includegraphics*[width=8.3 cm]{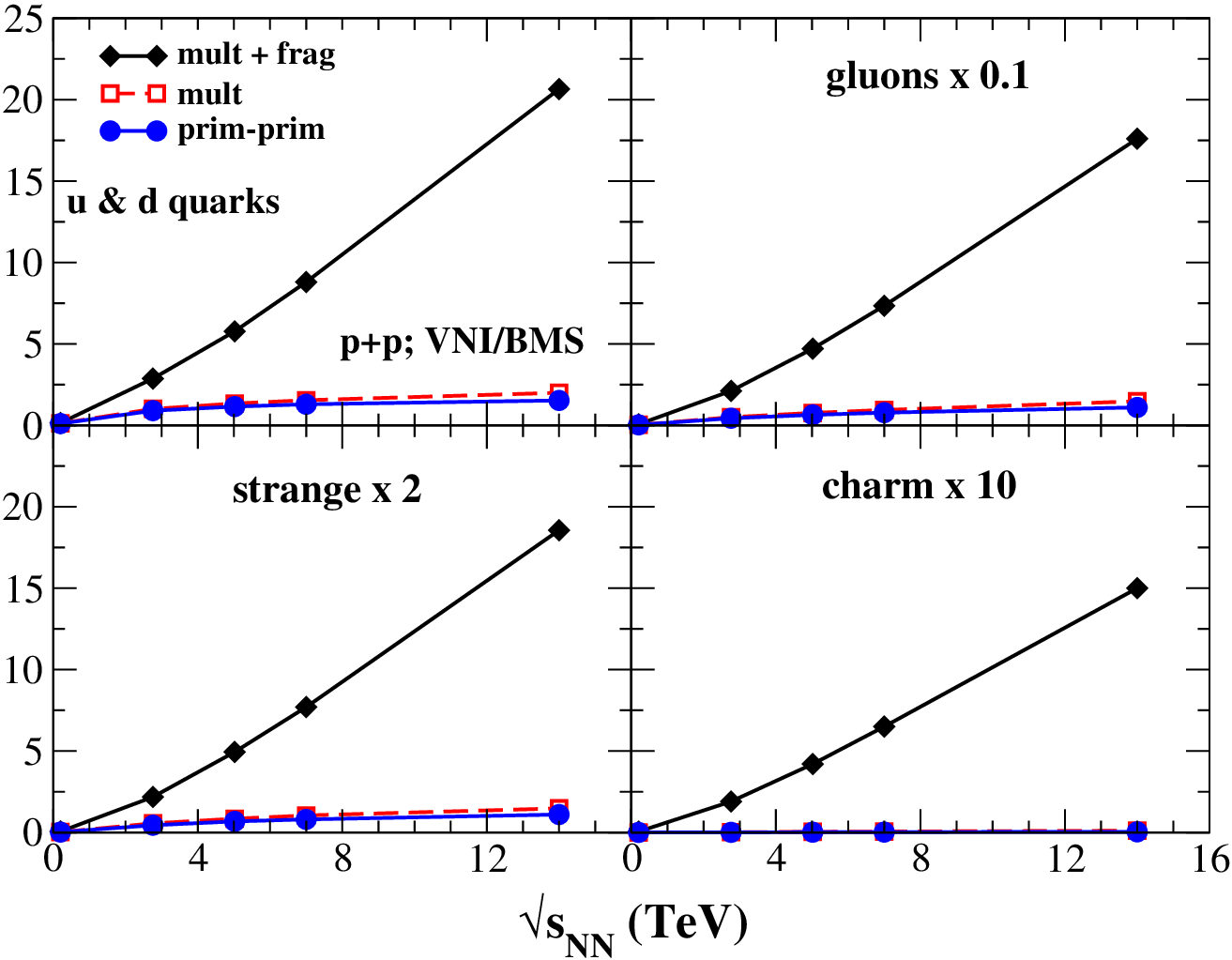}}
\centerline{\includegraphics*[width=8.3 cm]{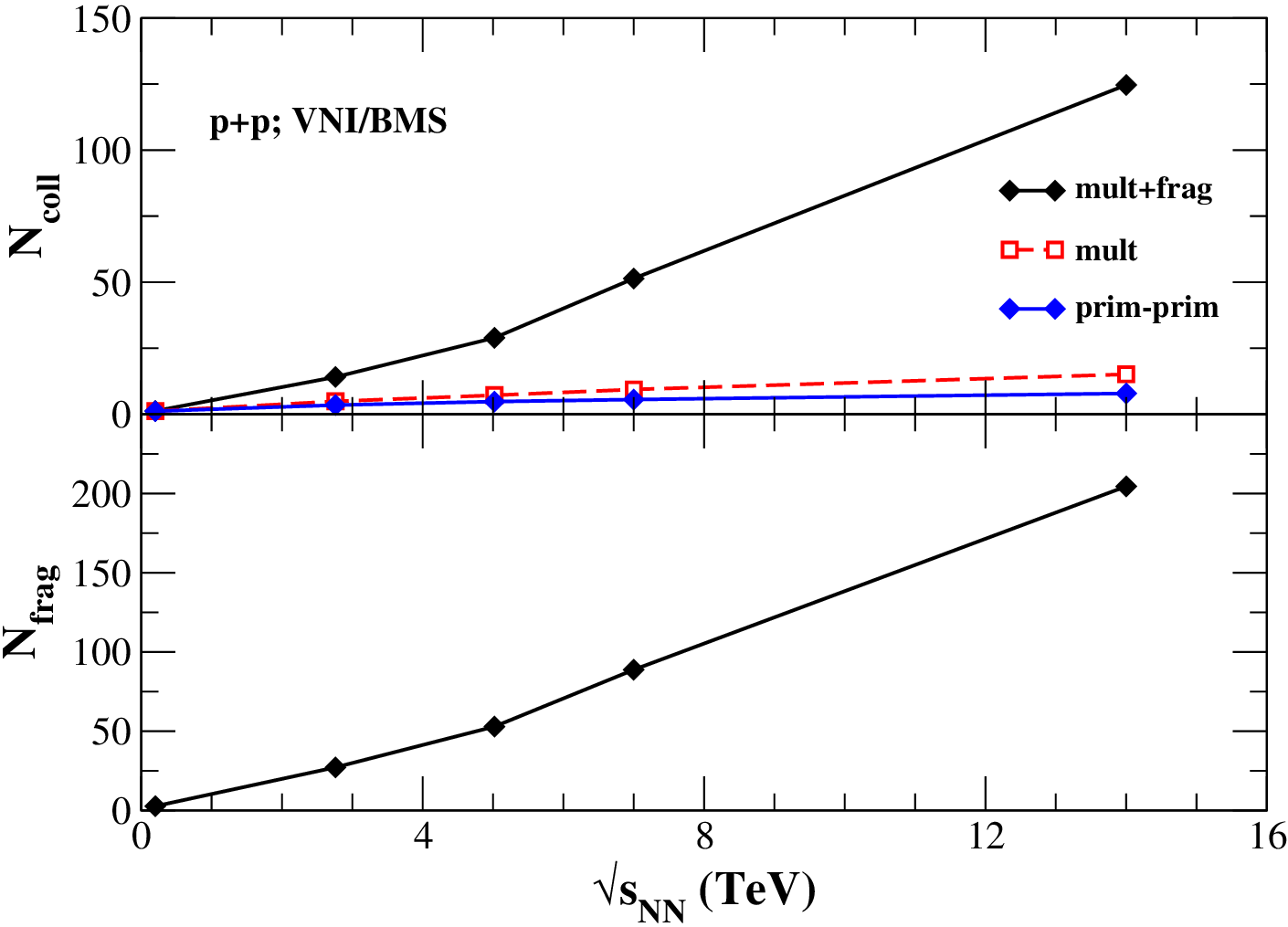}}
\caption{(Color online) [Upper panel] Production of light quarks, strange quarks, charm quarks and gluons
as a function of centre of mass energy for semi-hard partonic collisions in $pp$ system for 
calculations involving scattering only between primary partons (filled circles), multiple scatterings {\em with-out} fragmentation
 of scattered partons (hollow squares) and multiple scatterings {\em with} fragmentations of scattered partons (filled diamonds). [Lower panel] 
Number of collisions and number of fragmentations as a function of centre of mass energy.}
\label{sqrts}
\end{figure}

\section{Formulation} 

The Monte Carlo implementation of the parton cascade model
has been discussed in ~\cite{Geiger:1991nj,Bass:2002fh} for production of
light quarks, gluons, and heavy quarks.

The $2\rightarrow 2$ scatterings involving light quarks and gluons included in VNI/BMS are:
\begin{eqnarray} 
q_i q_j &\rightarrow& q_i q_j , \, q_i \bar{q}_i \rightarrow q_j \bar{q}_j \, , \nonumber\\
q_i\bar{q}_i &\rightarrow & gg , \, q_i \bar{q}_i \rightarrow g \gamma \, ,\nonumber\\
q_i \bar{q}_i &\rightarrow & \gamma \gamma , \, q_i g \rightarrow q_i g \, , \nonumber\\
q_i g & \rightarrow & q_i \gamma , \, gg \rightarrow q_i \bar{q}_i , \, \nonumber\\
gg &\rightarrow & gg ~.
\end{eqnarray}
The heavy quark production is  included~\cite{Younus:2013rja,Srivastava:2017bcm} via,
\begin{equation} 
q \bar{q} \rightarrow Q \bar{Q} , \, \, \, \, gg \rightarrow Q \bar{Q} \nonumber\\
\end{equation}
processes, while their scatterings with light quarks or gluons are included
via the 
\begin{equation}
q Q \rightarrow  qQ , \, \, \, g Q \rightarrow g Q ~,
\end{equation}
processes, where $g$ stands for gluons, $q$ stands for light quarks, 
and $Q$ stands for heavy quarks.

The $2\rightarrow 3$ reactions are included via time-like branchings of the final-state
partons (see Ref.~\cite{Bass:2002fh}) :
\begin{eqnarray}
g^{*} &\rightarrow&  q_i \bar{q}_i  \, , \, {q_i}^* \rightarrow q_i g \, , \nonumber\\
g^{*} &\rightarrow& gg \, , \, {q_i}^* \rightarrow q_i \gamma~,
\end{eqnarray}
following the well tested procedure adopted in PYTHIA.
\begin{figure}
\centerline{\includegraphics*[width=8.0 cm]{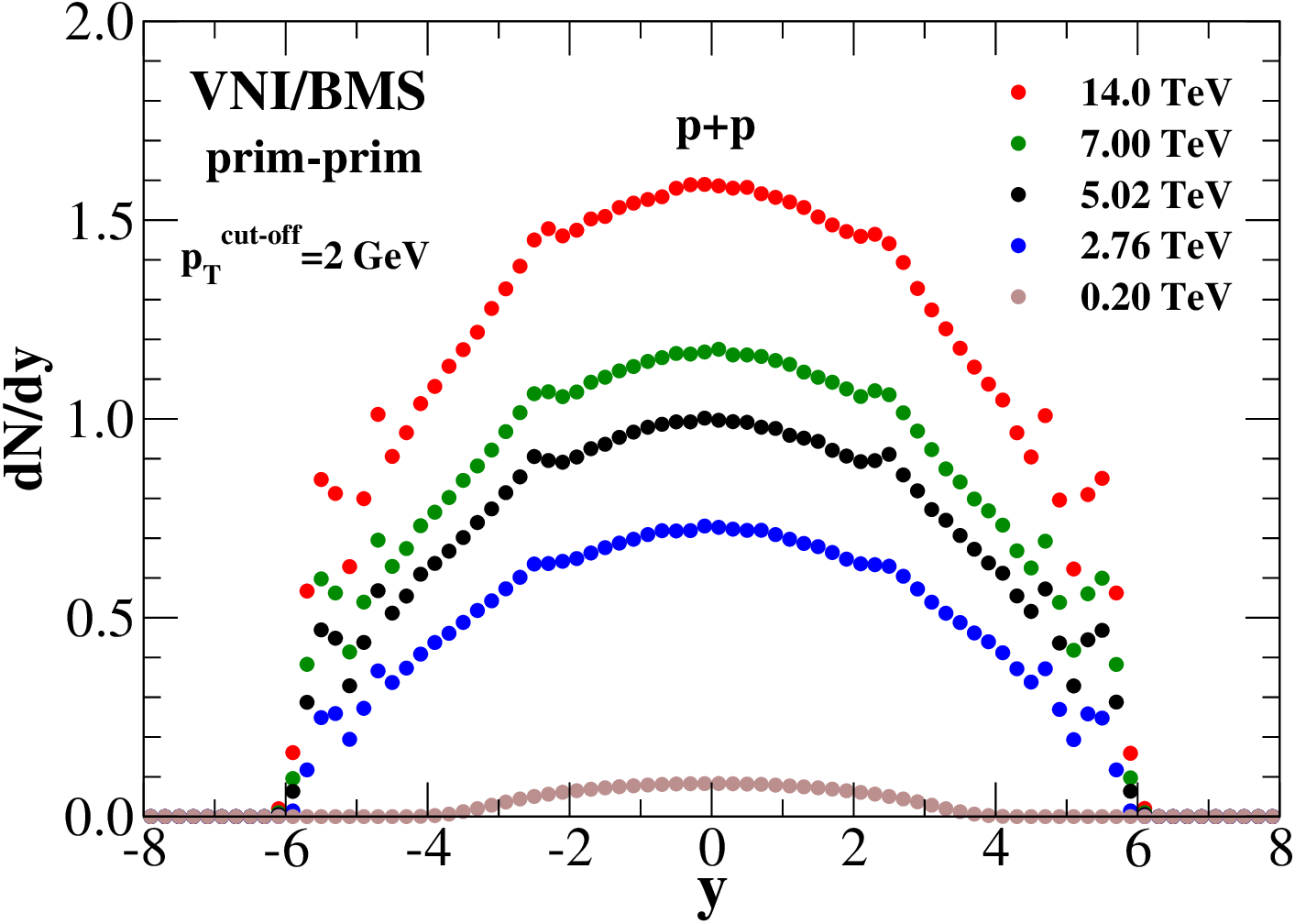}}
\centerline{\includegraphics*[width=8.0 cm]{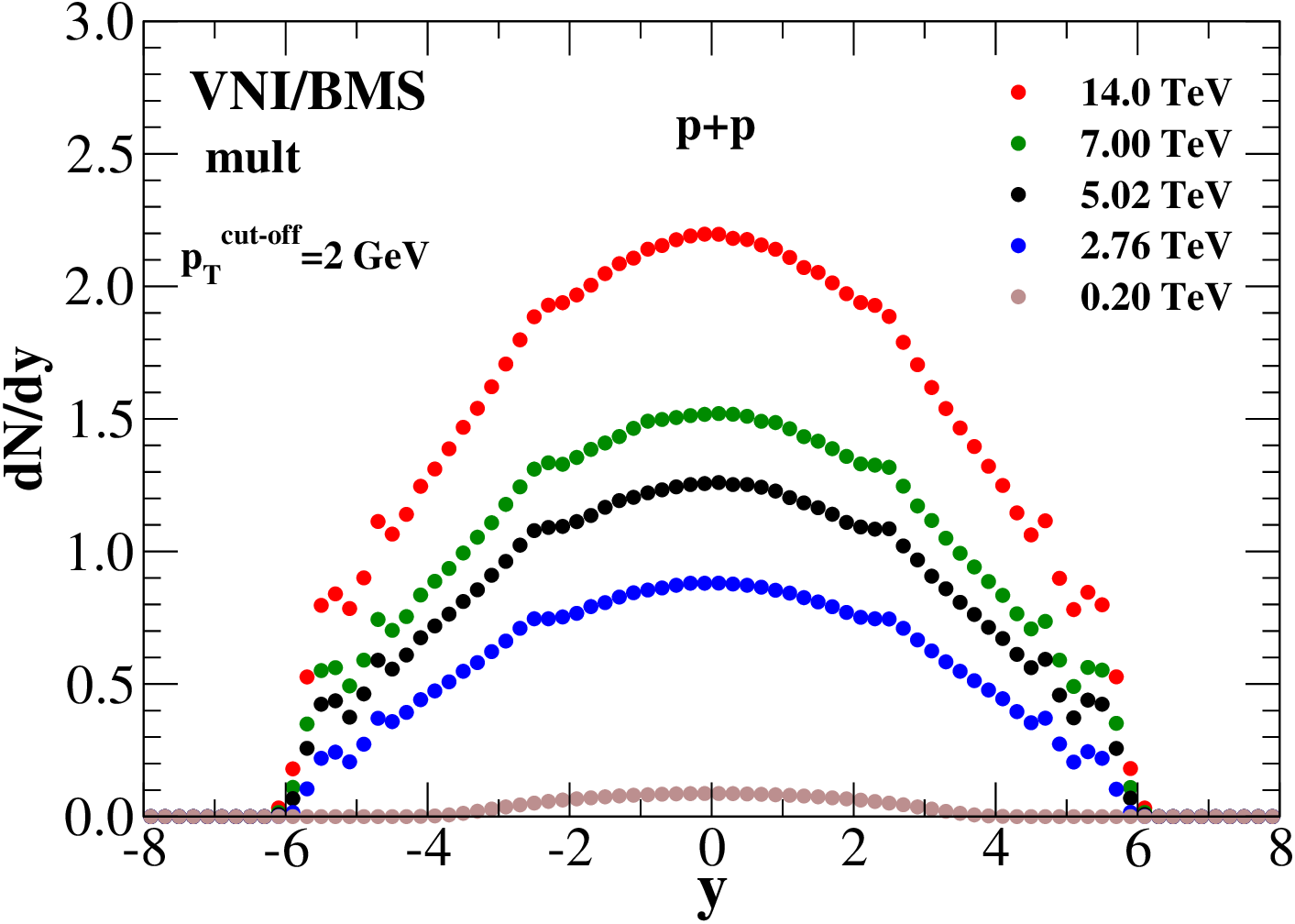}}
\centerline{\includegraphics*[width=8.0 cm]{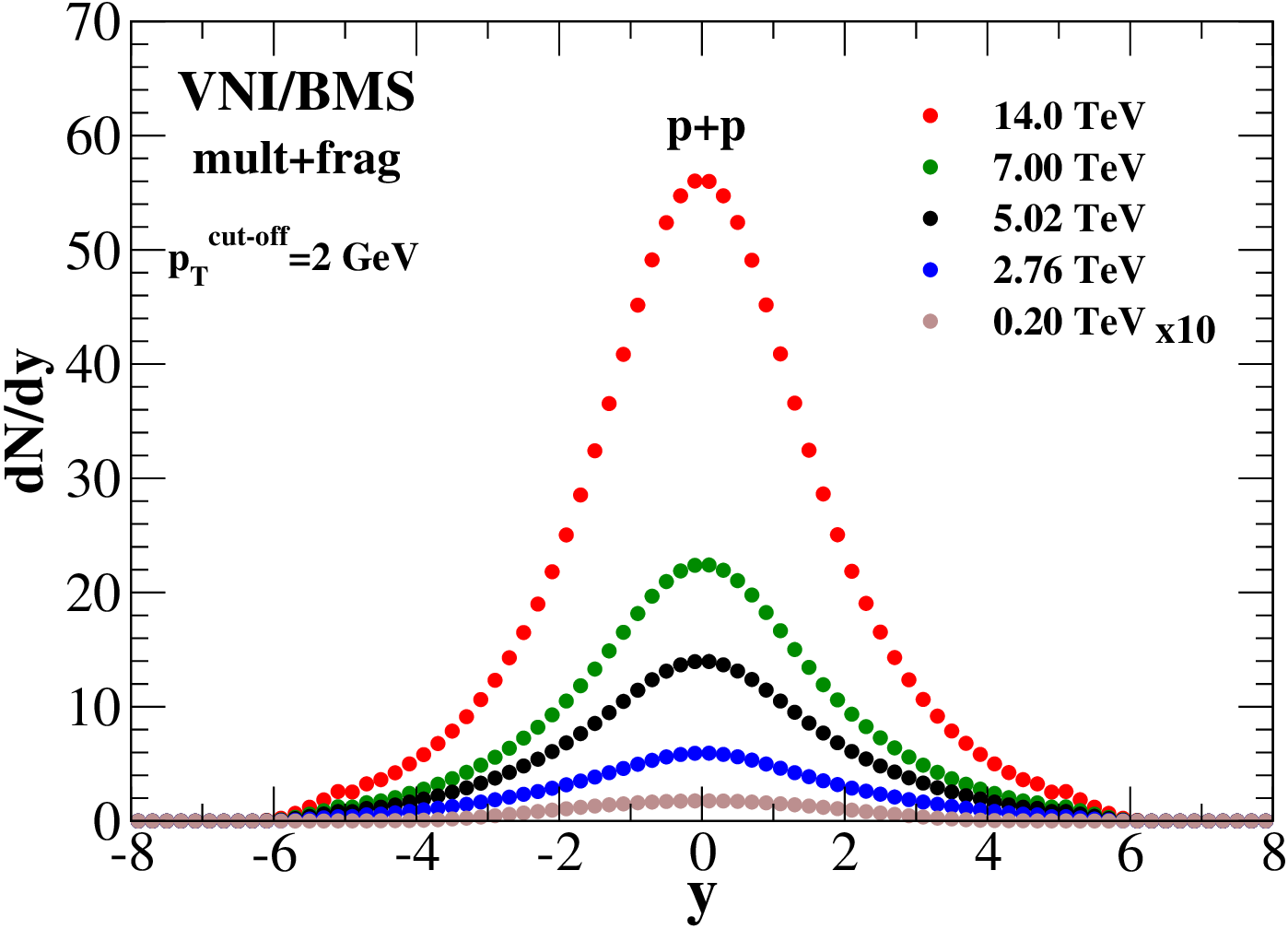}}
\caption{(Color online) Rapidity density of partons produced in $pp$ interactions
due to semi-hard collisions among primary partons (upper panel), multiple collisions
 {\em without} fragmentations of final state partons (middle
panel) and multiple collisions {\em with} fragmentations of final state
partons (lower panel) at different centre of mass energies.}  
\label{dndy-sqrts}
\end{figure}

\begin{figure}
\centerline{\includegraphics*[width=8.0 cm]{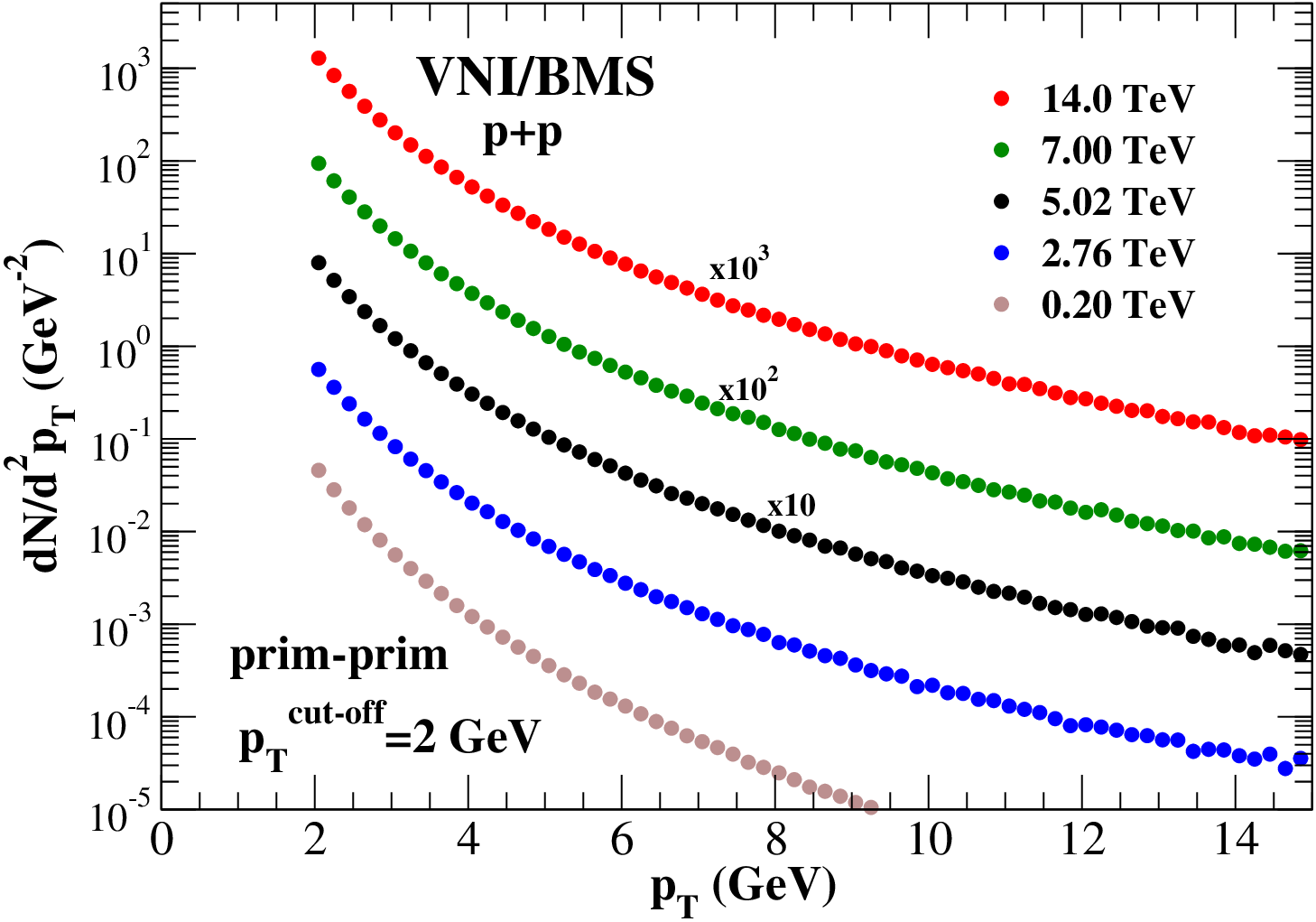}}
\centerline{\includegraphics*[width=8.0 cm]{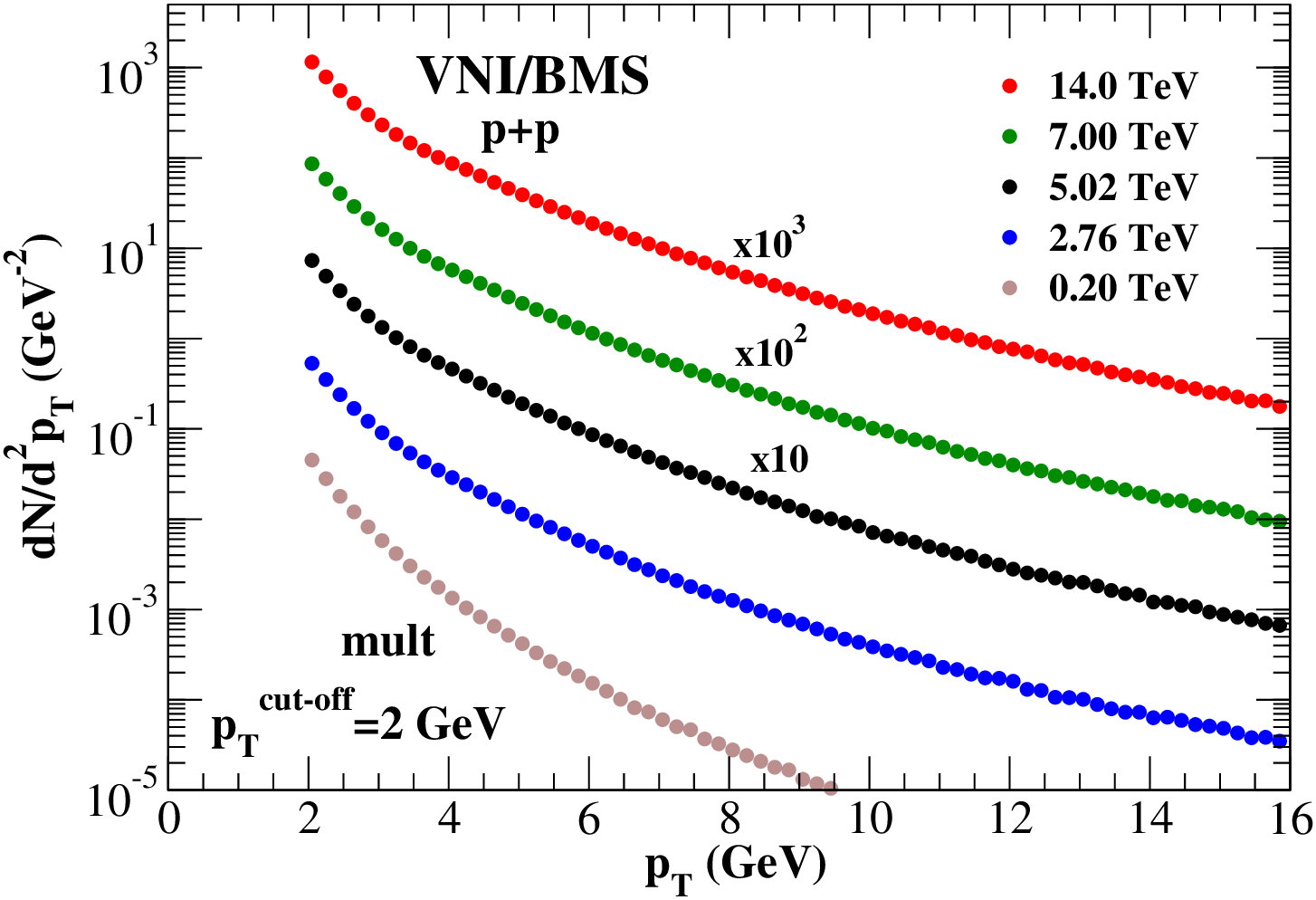}}
\centerline{\includegraphics*[width=8.0 cm]{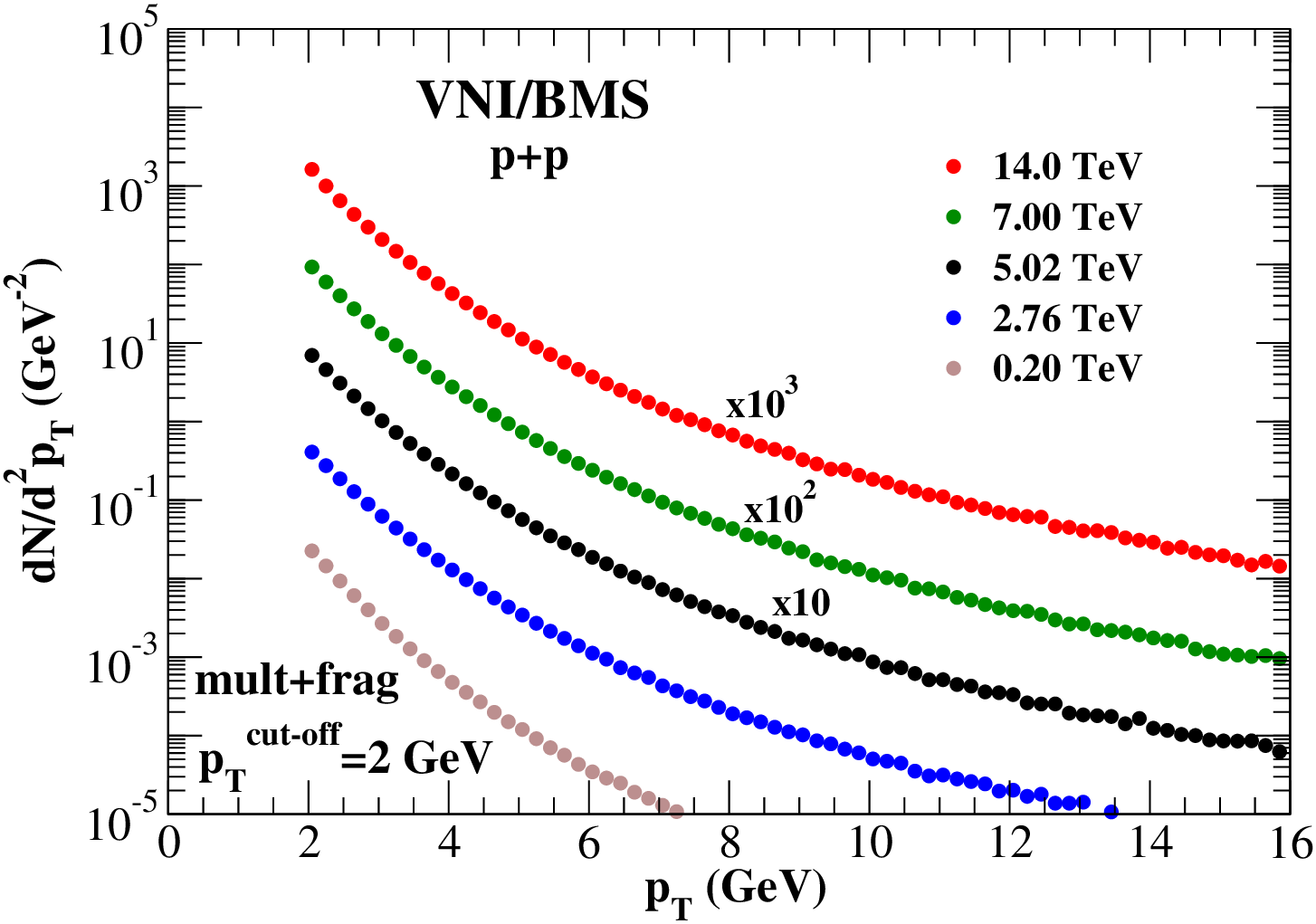}}
\caption{(Color online) Rapidity integrated $p_T$ spectra of partons produced in $pp$ interactions
due to semi-hard collisions among primary partons (upper panel), multiple collisions
 {\em without} fragmentations of final state partons (middle
panel) and multiple collisions {\em with} fragmentations of final state
partons (lower panel) at different centre of mass energies.}  
\label{spec-sqrts}
\end{figure}
We add that the IR-singularities in these pQCD cross-sections are avoided in PCM
by introducing a lower cut-off on the momentum transfer $p_T^\text{cut-off} \approx $ 2 GeV.
We also add that, as discussed in Ref.~\cite{Srivastava:2017bcm}, the processes $gQ \rightarrow gQ$ and
$qQ \rightarrow qQ$ have been explicitly excluded in these calculations when the heavy quark
belongs to the sea, in order to account for the strong suppression of these interactions 
when NLO terms are included.

\begin{figure}
\centerline{\includegraphics*[width=8.0 cm]{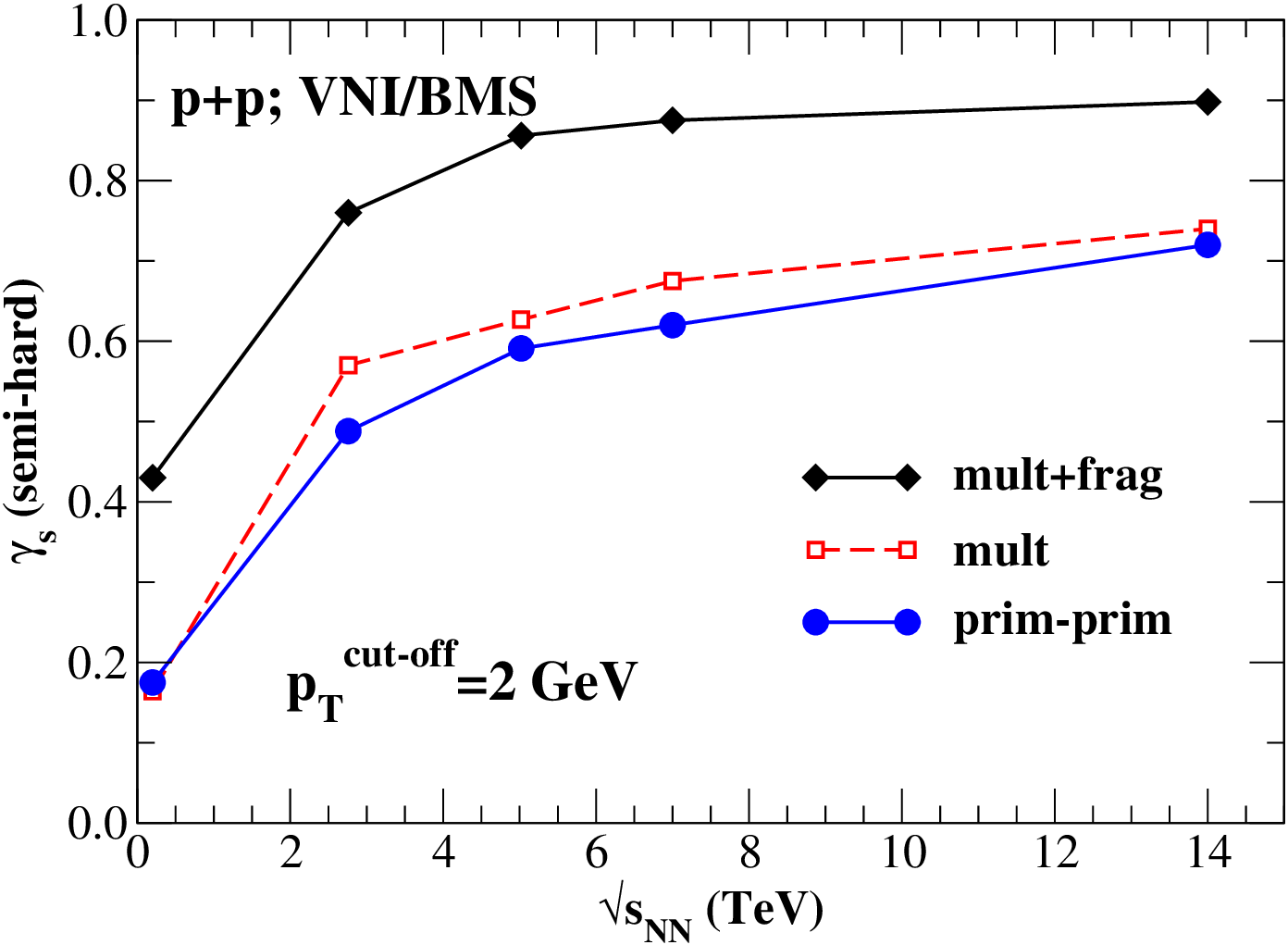}}
\caption{(Color online) The ratio of strange quarks and light quarks produced in 
semi-hard processes for the three set of calculations discussed here.} 
\label{gamma-sqrts}
\end{figure}
The $2 \rightarrow 3$ processes are included by inclusion of radiative processes for the
 final state partons in a leading logarithmic approximation. The collinear sigularities 
have been regularized by terminating the time-line branchings, once the virtuality of the parton
drops to $Q_0^2=m_i^2+\mu_0^2$, where $m_i$ is the current mass of the parton (zero for gluons,
current mass for quarks) and $\mu_0 $ has been kept fixed as 1 GeV. 
We have included $g \rightarrow gg$, $q \rightarrow q g$, 
$g \rightarrow q \bar{q}$, and $q \rightarrow q \gamma$ branchings for 
which the relevant branching functions
$P_{a\rightarrow bc}$ are taken from Altarelli and Parisi~\cite{Altarelli:1977zs}. 
\begin{figure}
\centerline{\includegraphics*[width=8.0 cm]{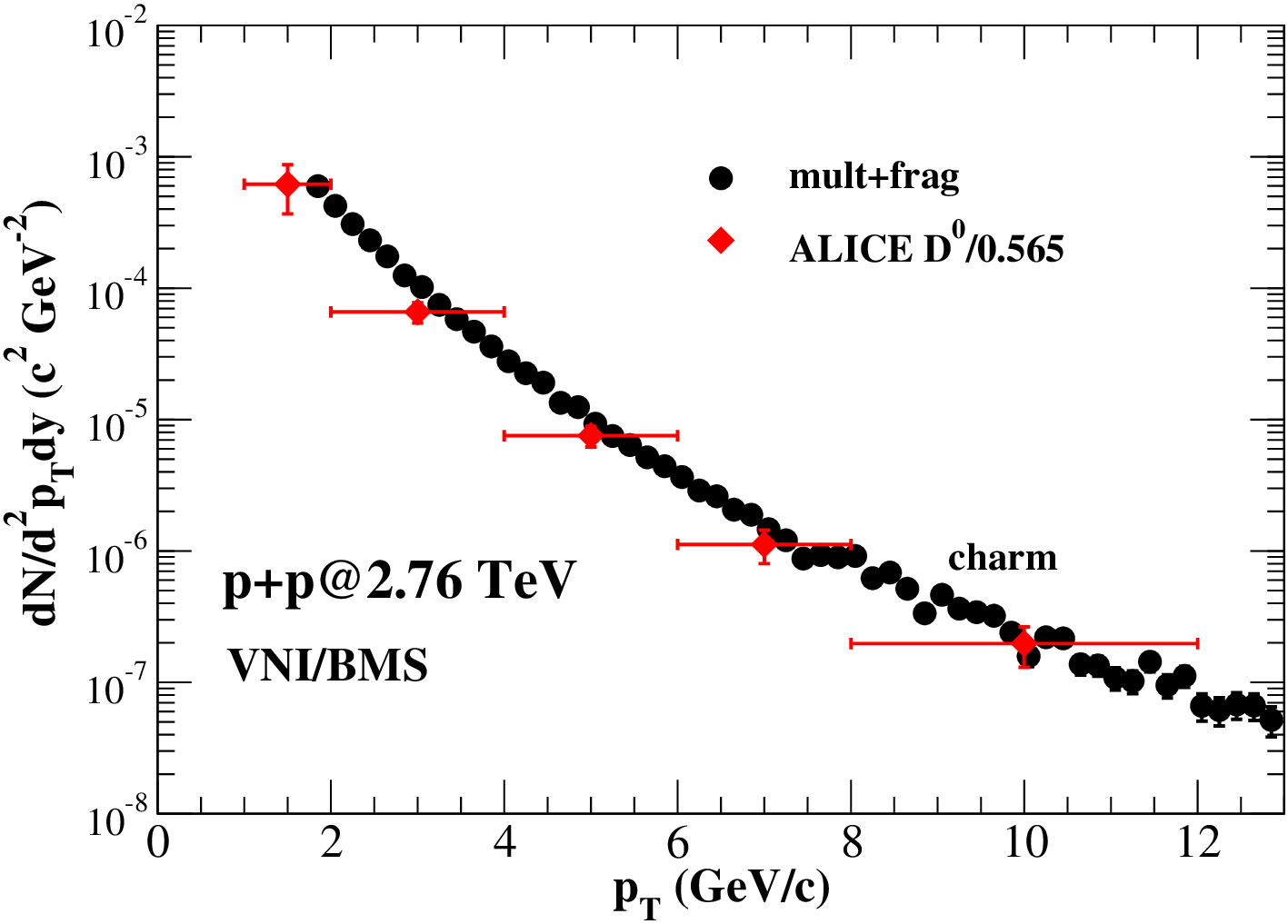}}
\caption{(Color online) A comparison of our calculations with prompt charm production
measured by ALICE experiment in $pp$ collisions at 2.76 TeV~\cite{Abelev:2012vra}.}
\label{2.76_data}
\end{figure}
\begin{figure}
\centerline{\includegraphics*[width=8.0 cm]{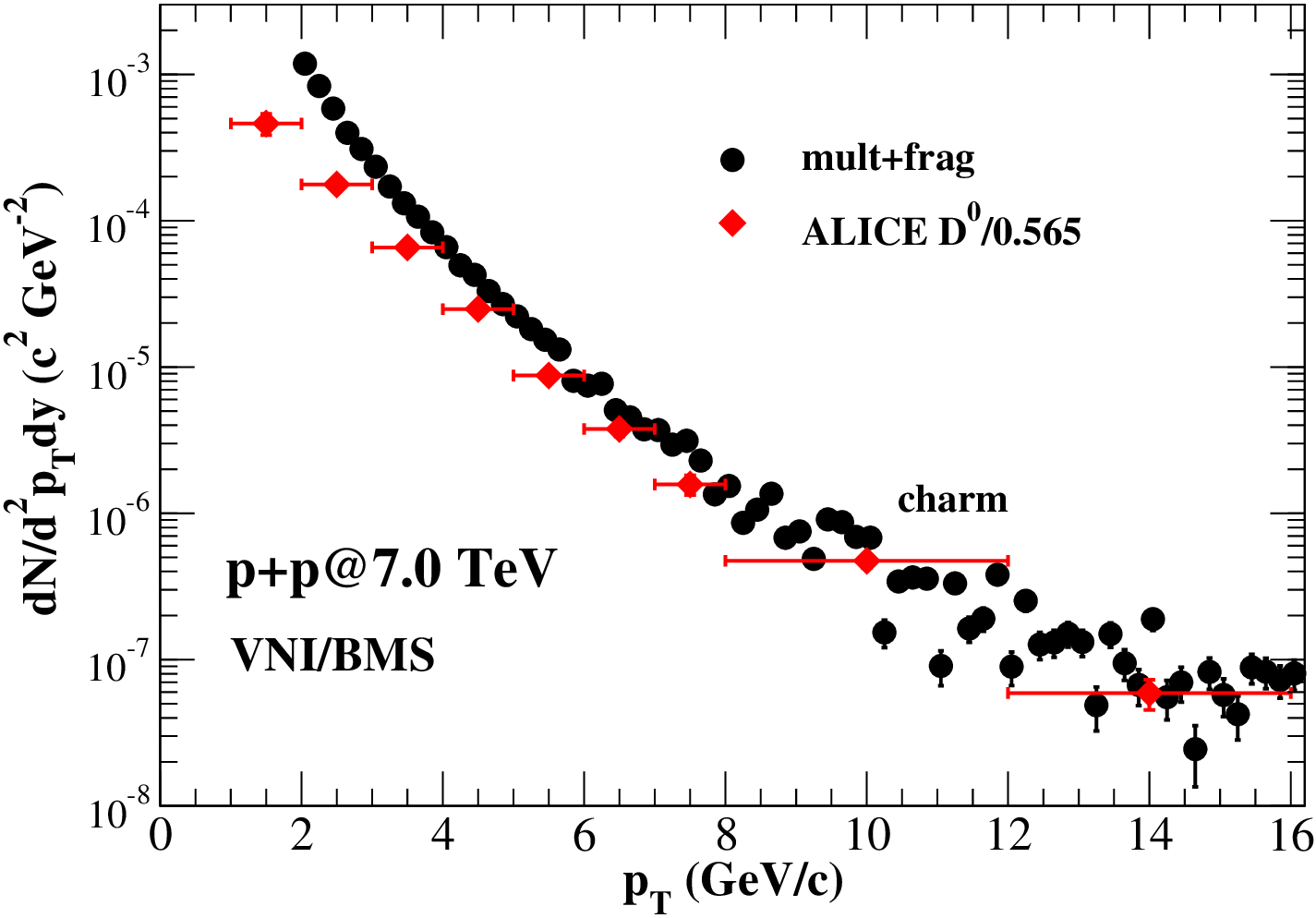}}
\caption{(Color online) A comparison of our calculations with prompt charm production
measured by ALICE experiment in $pp$ collisions at 7.00 TeV~\cite{ALICE:2011aa}.}
\label{7.00_data}
\end{figure}
The initial state of the nucleons has been set up in terms of partons whose momentum distributions 
are described by the parton distribution functions initialized at the scale of $Q_\text{ini}^2$ = 4 GeV$^2$. 
We have used GRV-HO function for our studies
even though more modern functions are now available in the literature, primarily as we are more interested
in the evolution of the multi-parton interactions when the centre of mass energy or 
the impact parameter or the lower momentum cut-off for parton scattering is altered.
The partons are distributed around the centres of nucleons according to the distribution,
\begin{equation}
h_N(\vec{r})= \frac{1}{4\pi} \frac{\nu^3}{8\pi}\exp(-\nu r)
\label{nucleon}
\end{equation}
with $\nu$ chosen to give the root mean square radius $R_N^\text{ms}\equiv\sqrt{12/\nu}= $ 0.81 fm.

Unless otherwise stated, we have kept $p_T^\text{cut-off}$ fixed at 2 GeV. Most of our studies 
using PCM at RHIC energies used a more modest value of $p_T^\text{cut-off} \approx$ 0.78 GeV.
For the corresponding results at $\sqrt{s}$ =0.2 TeV, the reader is referred to Ref.~\cite{Bass:2002fh,Bass:2002vm,Bass:2003mk,Chang:2004eha,Renk:2005yg}. Our calculations do not consider hadronization and subsequent interactions among the hadrons. 

\section{Analysis Setup}

As mentioned earlier, we perform three sets of calculations to investigate the
essential features of the evolution of the partonic cascade in $pp$ collisions. 
We define primary partons as partons which constitute the nucleon and which have not undergone
any interaction. The secondary partons are those which are produced in collisions or fragmentation
of scattered partons.

The first set of calculations look at the system which would be formed if only
primary-primary collisions are included in the calculations.
The second set of calculations look at the system when primary-primary, primary-secondary,
and secondary-secondary collisions are permitted but fragmentation of final state partons
is not permitted, thus effectively blocking parton multiplication.
The final set of calculations describe the system when all possible multiple scatterings
among partons are tracked and when the final state partons fragment, leading to
a substantial increase in number of collisions and parton production from semi-hard processes.

We discuss our results in terms of partons produced in these semi-hard interactions and number of collisions as well
as number of fragmentations (when applicable). We also give our results for relative abundance of strange quarks 
with respect to light quarks which are produced by
semi-hard interactions considered here, defined by:
\begin{equation}
\gamma_s^{\rm{semi-hard}}  = \frac {2( N_s + N_{\bar s})} {N_u + N_{\bar u} + N_d + N_{\bar d}} \, .
\label{gamma}
\end{equation}
 \begin{figure}
\centerline{\includegraphics*[width=8.3 cm]{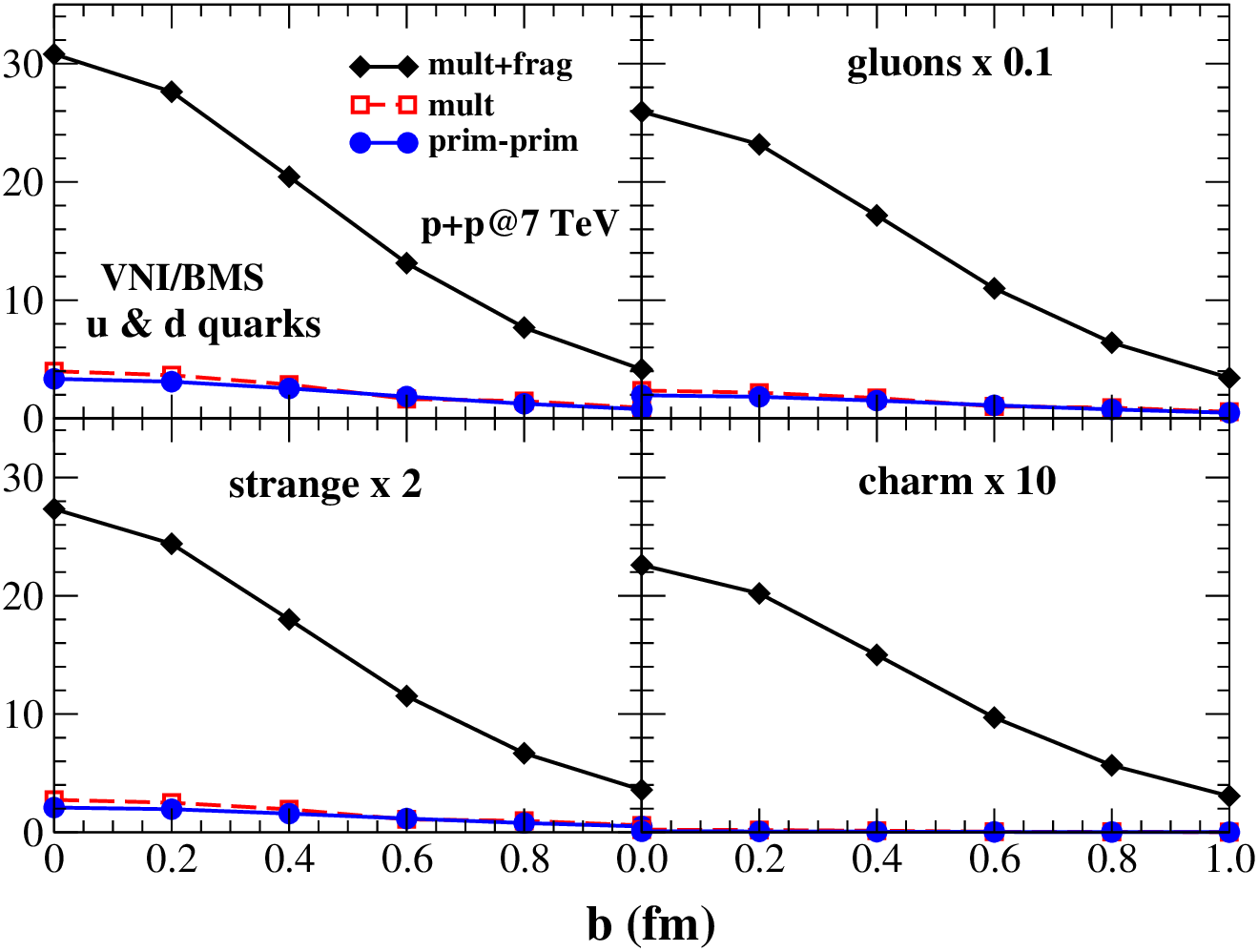}}
\centerline{\includegraphics*[width=8.3 cm]{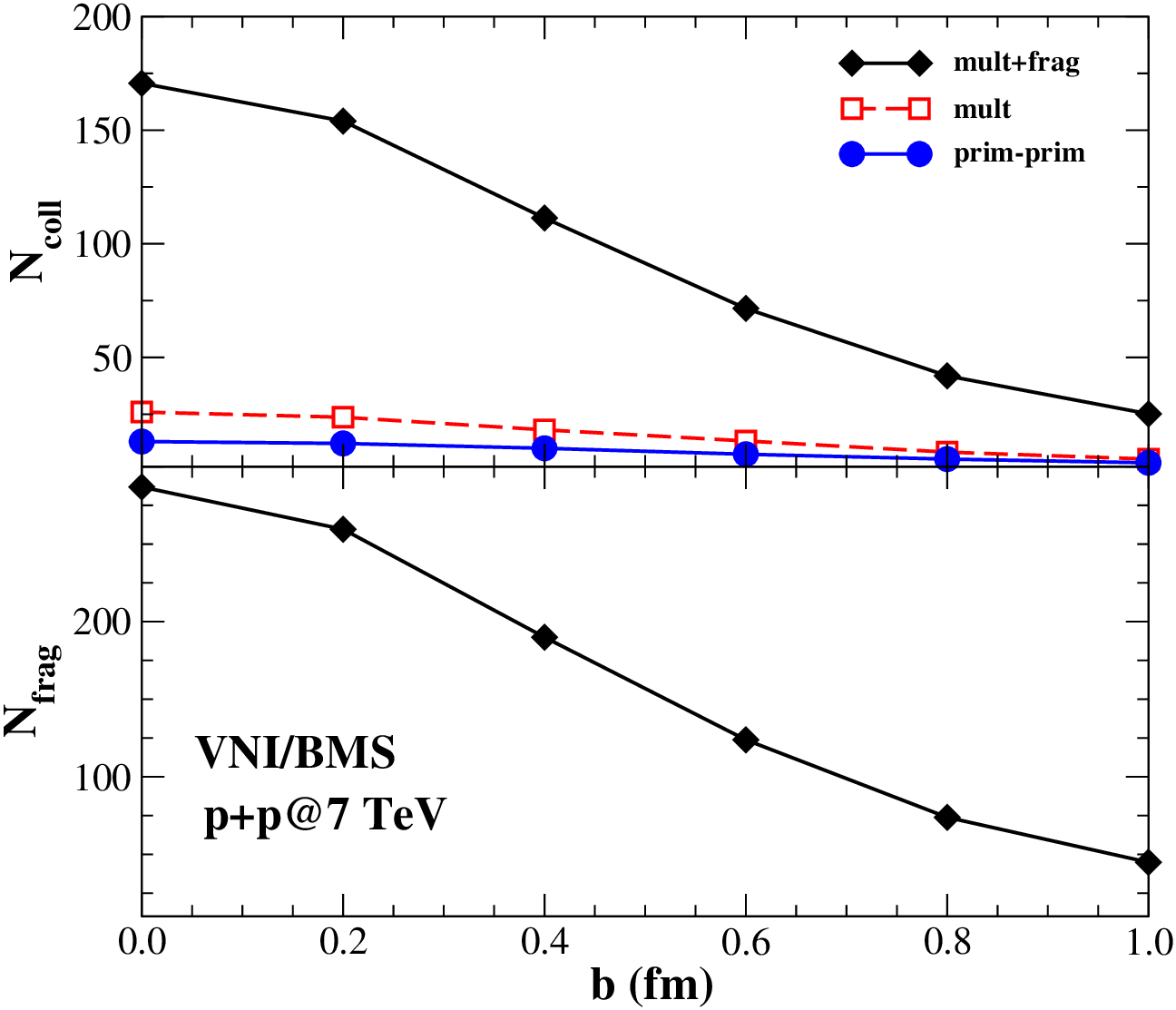}}
\caption{(Color online) [Upper panel] Production of light quarks, strange quarks, charm quarks and gluons 
as a function of impact parameter for semi-hard partonic collisions in $pp$ system for 
calculations involving
scattering only between primary partons (filled circles), multiple scatterings {\em with-out} fragmentation
 of scattered partons (hollow squares)
and multiple scatterings {\em with} fragmentation of scattered partons (filled diamonds). [Lower panel] Number of collisions and number of fragmentations as a function of impact parameter.}
\label{impact}
\end{figure}
We re-emphasize that the multi-parton interactions included in these calculations 
take place only if the momentum-transfer is larger than the $p_T^\text{cut-off} \approx$ 2 GeV.
This necessarily provides that quite a large part of the initial state partons continue without interaction.
These include valence/sea light quarks, sea strange quarks and gluons, which move with the
momenta with which they were initialized. Many more partons would interact if $p_T^\text{cut-off}$ 
is lowered or if a suitably screened interaction, e.g, by including Debye screening is considered\footnote{The use of a Debye screening mass poses conceptional challenges since it implies the formation of a thermalized medium}. The lack of interactions below the $p_T^\text{cut-off}$ implies that we cannot directly study the possible formation of a thermalized medium, which would require abundant interactions at scales below the $p_T^\text{cut-off}$. However, we can ascertain whether a sufficient number of interacting partons is deposited into the system that could potentially lead to the formation of a thermalized medium.

\begin{figure}
\centerline{\includegraphics*[width=8.0 cm]{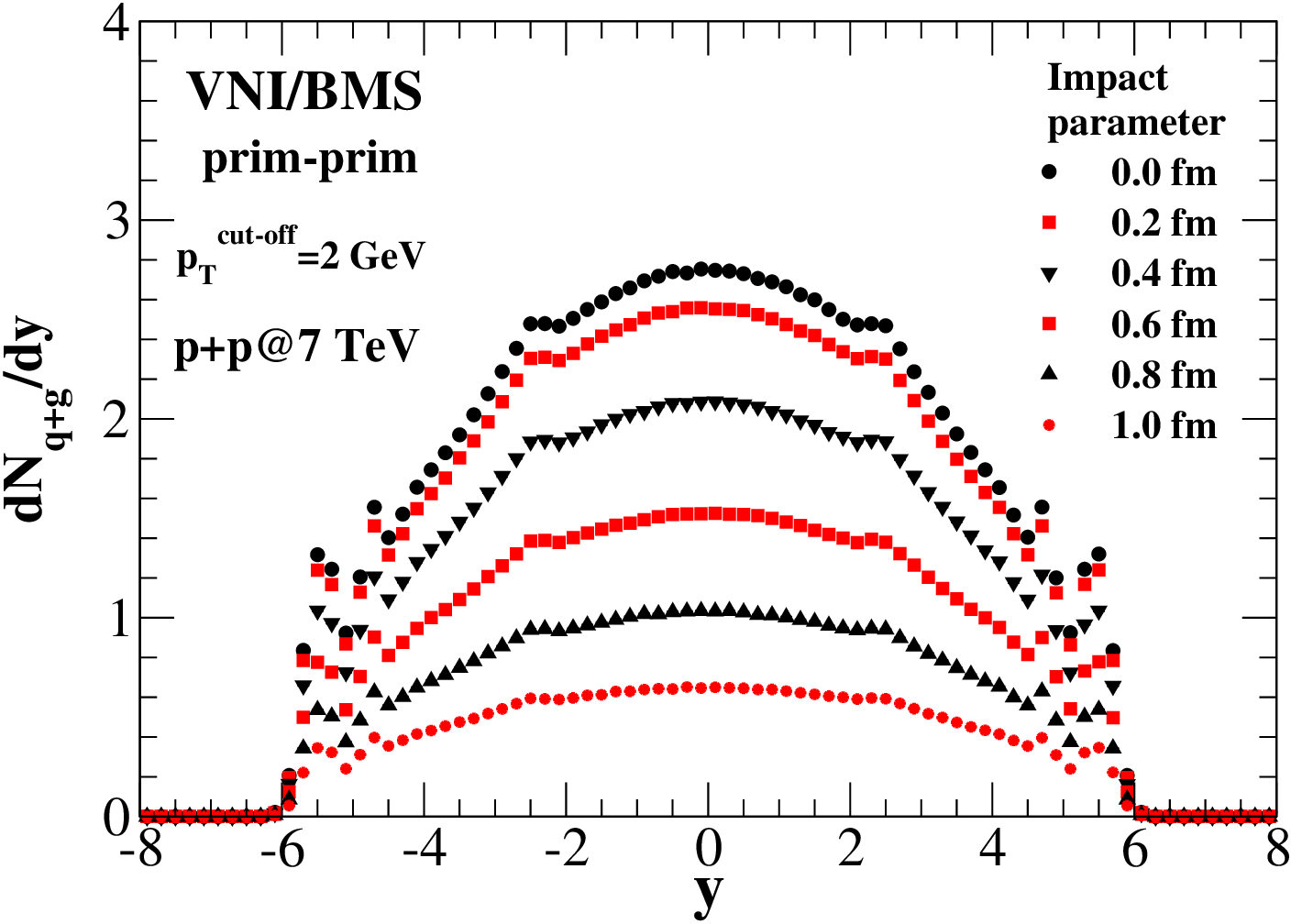}}
\centerline{\includegraphics*[width=8.0 cm]{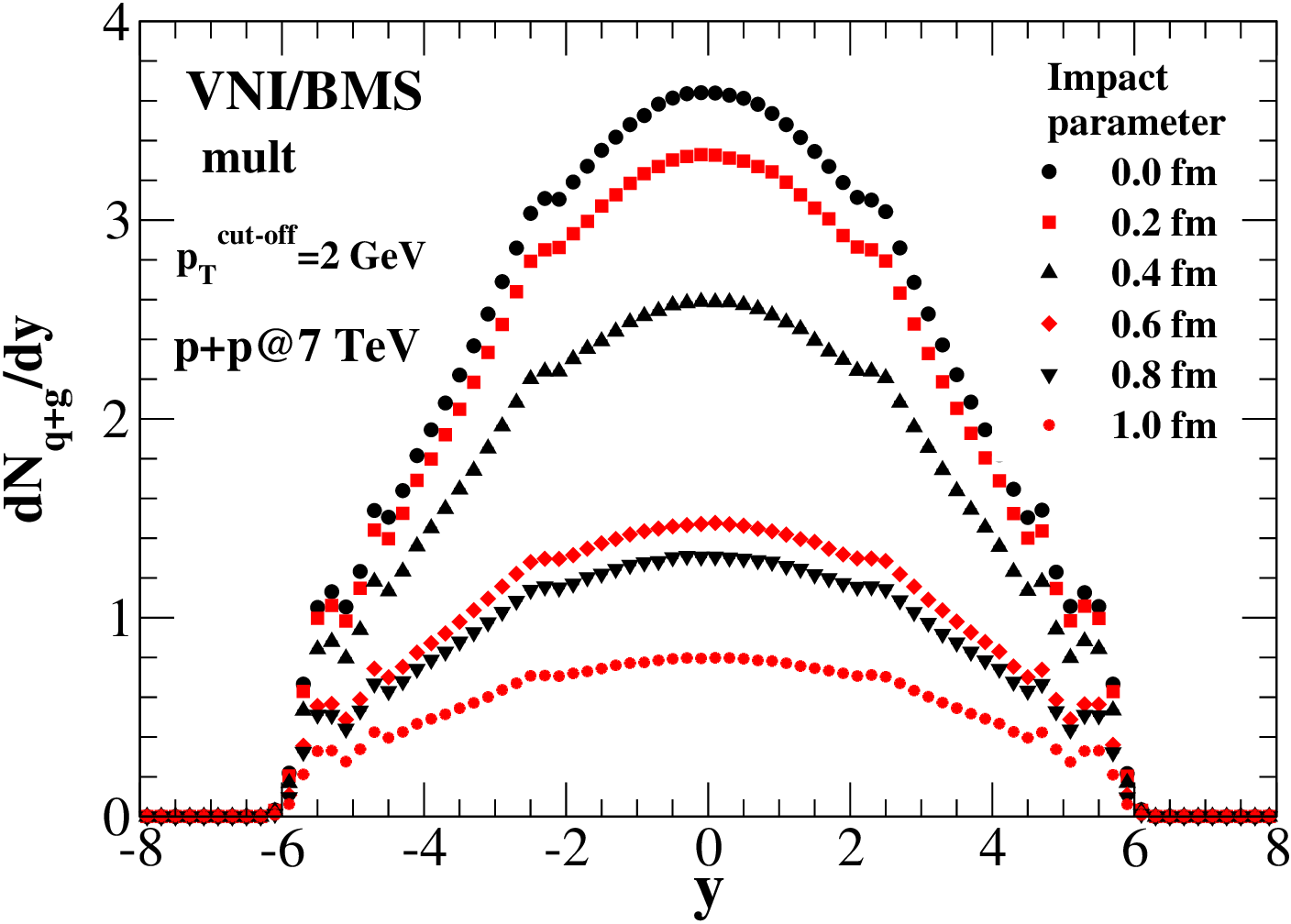}}
\centerline{\includegraphics*[width=8.0 cm]{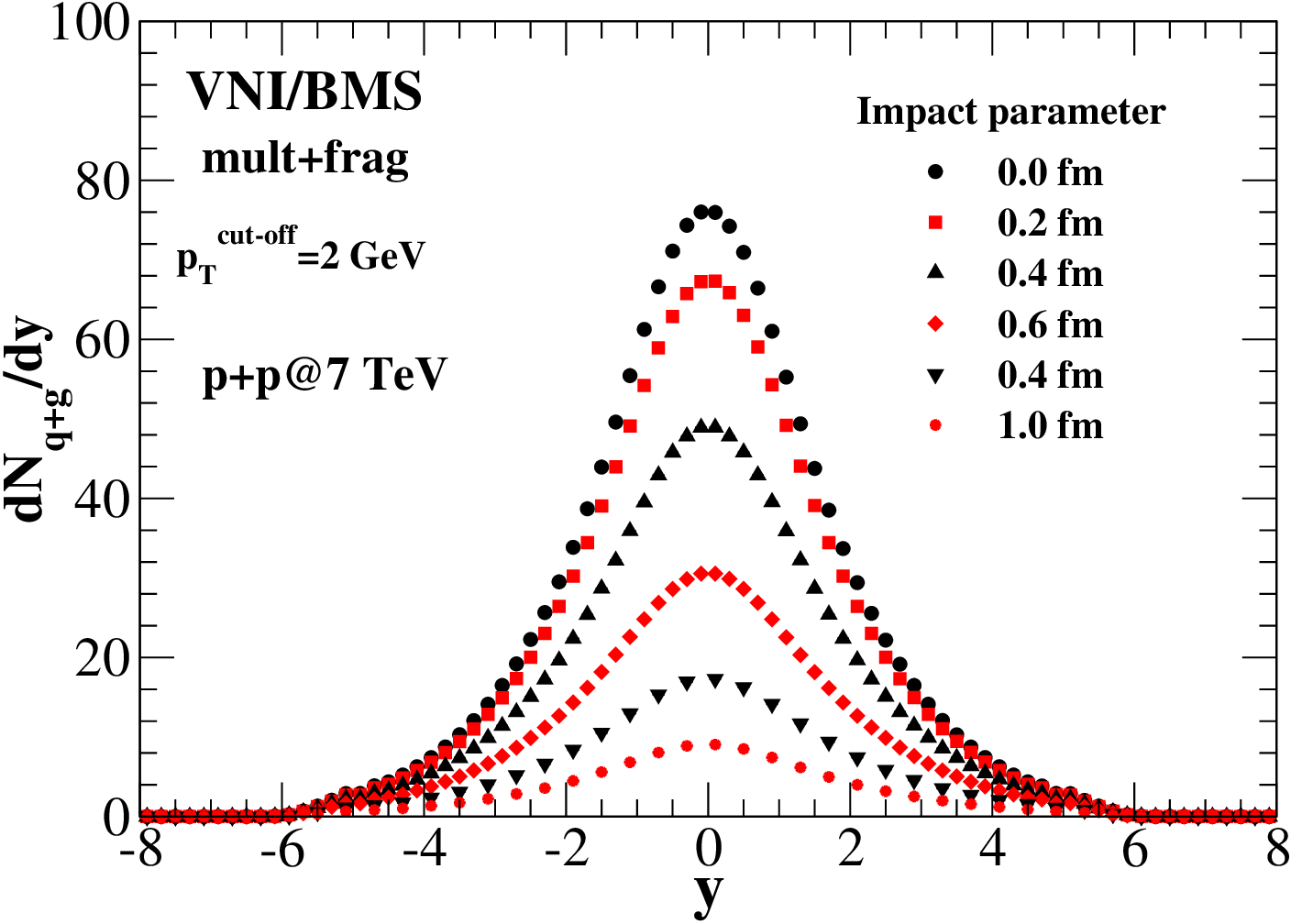}}
\caption{(Color online) Rapidity density of partons produced in $pp$ interactions at
$\sqrt{s}=$ 7 TeV
due to semi-hard collisions among primary partons (upper panel), multiple collisions
 {\em without} fragmentations of final state partons (middle
panel) and multiple collisions {\em with} fragmentations of final state
partons (lower panel) at different impact parameters.}  
\label{dndy-b}
\end{figure}

\begin{figure}
\centerline{\includegraphics*[width=8.0 cm]{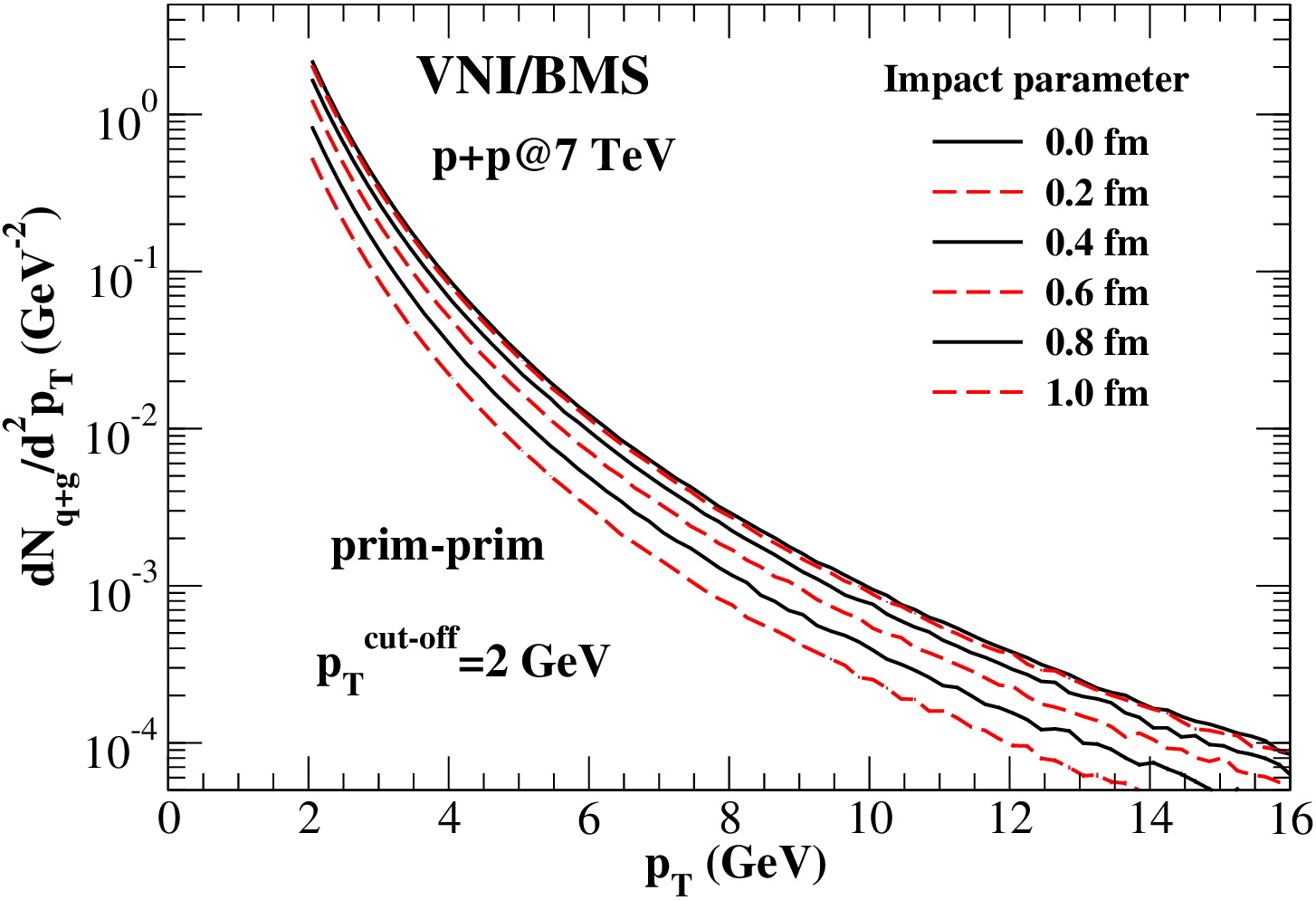}}
\centerline{\includegraphics*[width=8.0 cm]{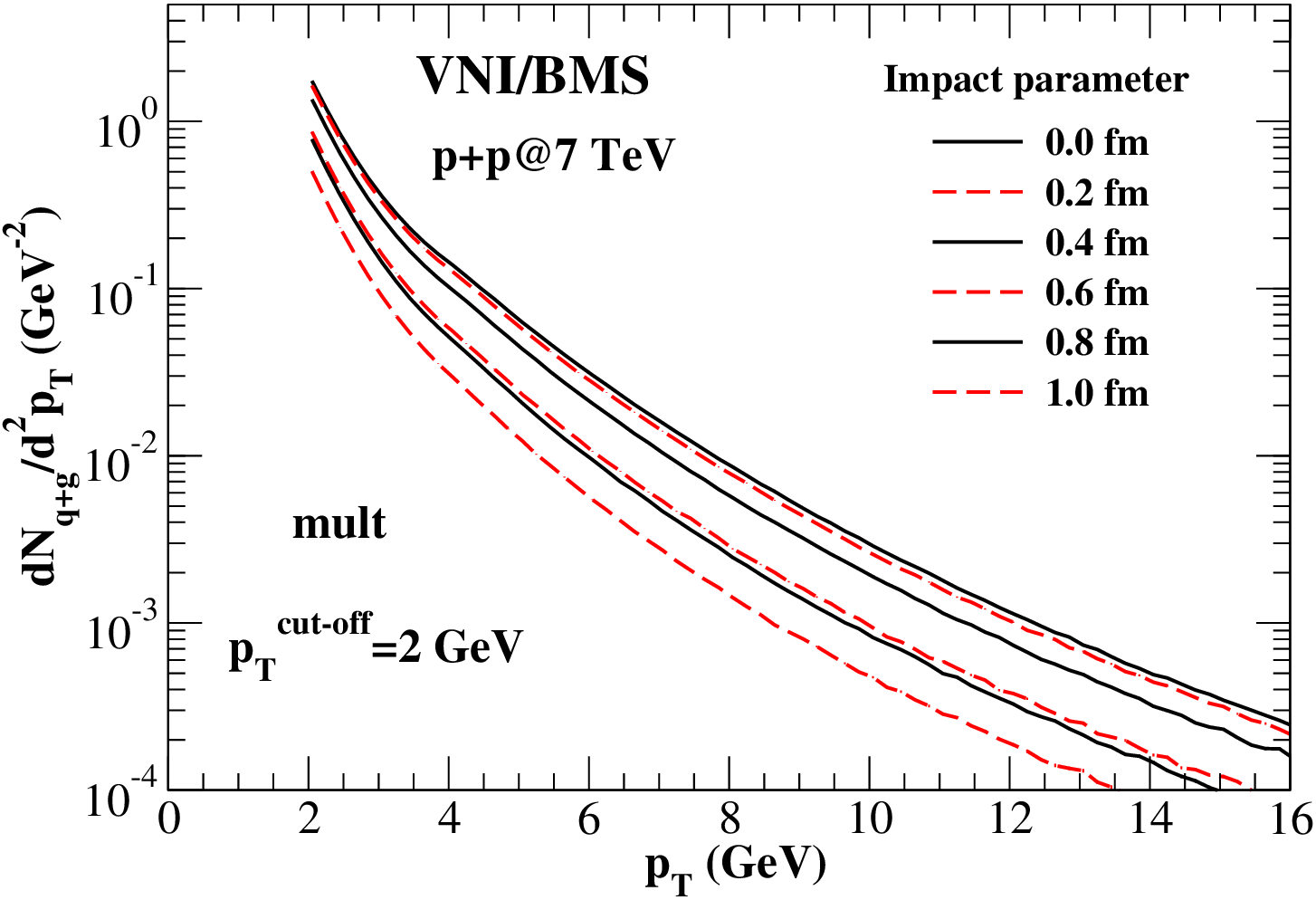}}
\centerline{\includegraphics*[width=8.0 cm]{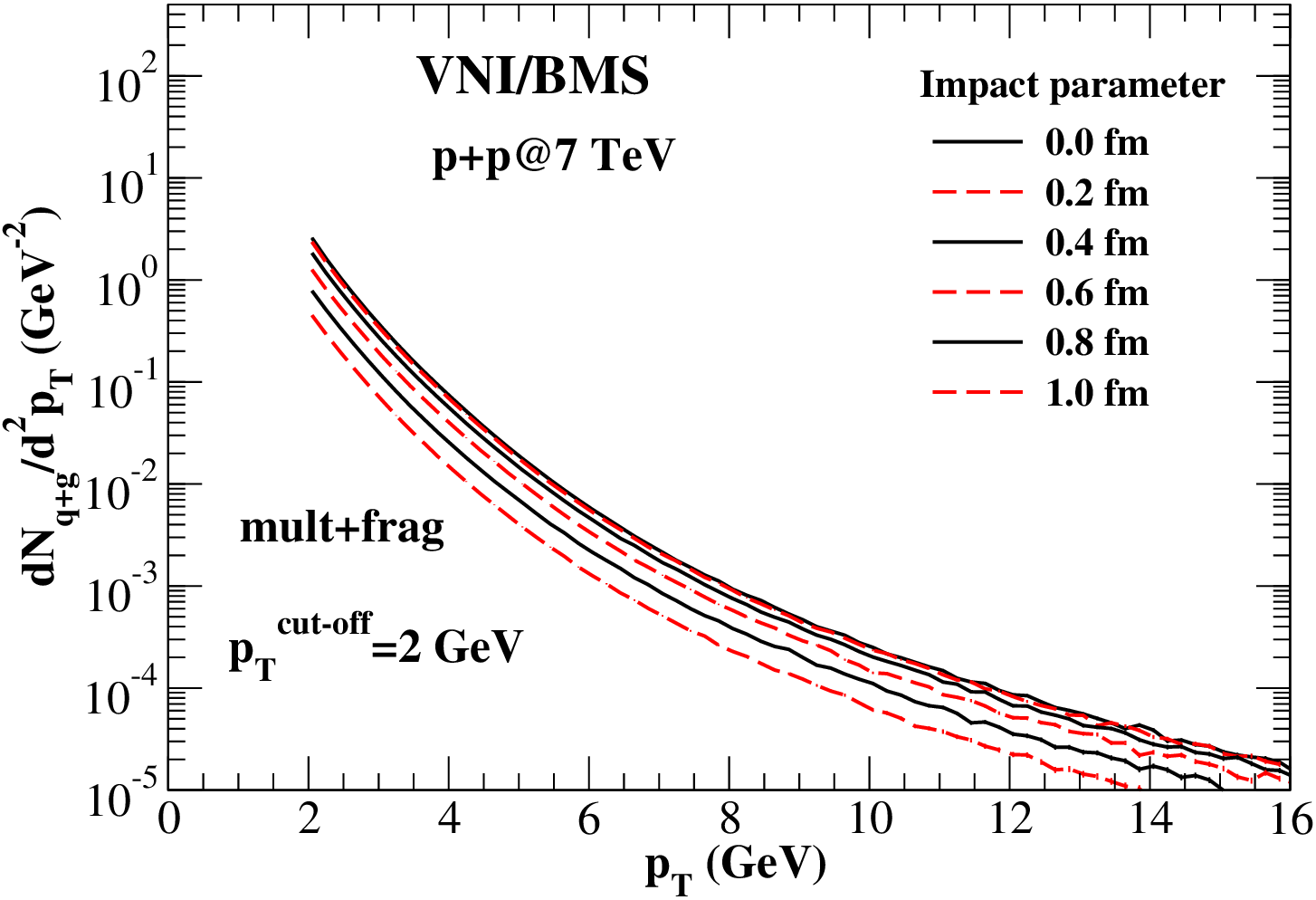}}
\caption{(Color online) Rapidity integrated $p_T$ spectra of partons produced in $pp$ interactions
at $\sqrt{s}$ = 7 TeV
due to semi-hard collisions among primary partons (upper panel), multiple collisions
 {\em without} fragmentations of final state partons (middle
panel) and multiple collisions {\em with} fragmentations of final state
partons (lower panel) at different impact parameters.}  
\label{spec-b}
\end{figure}

\section{Evolution of multi-parton interactions with centre of mass energy}


We first look at partons produced in minimum bias collisions of protons at varying
center of mass energies (Fig.~\ref{sqrts}). We see only a marginal difference between
the results for the calculations involving primary-primary collisions and multiple collisions
(without fragmentation). The production of gluons, light quarks,
and strange quarks is seen to rise monotonically along with the number of 
semi-hard collisions with the center of mass energy of the $pp$ collision, but overall the number of partons involved in these interactions remains small, even at the highest beam energies. 
 
These results imply that semi-hard $2 \rightarrow 2$ interactions,
without fragmentation of final state partons, are too few in $pp$
collisions even at the highest energy considered  to lead to a
hot and dense interacting medium. 

The corresponding results for the calculations with fragmentation of final state
partons using the procedure indicated earlier, show a rapid multiplication 
of partons and a sharp rise in the number of collisions compared to the cases 
discussed above. This increase is clearly driven by the number of fragmentations,
We also see a sharp increase in production of strange and charm quarks due to the
multiple interactions as gluons multiply and interact. The increase could also be partly due  to the opening up of
processes like $g^{*} \rightarrow Q \overline{Q}$ as the centre of mass energy increases. 
%
%
We note that already at $\sqrt{s}_{NN}=200$ GeV more than 100 interacting partons are deposited into the system, which can participate in the partonic collisions having momentum transfers of 
more than 2 GeV. This number grows to about 500 at 2.76 TeV. 
These should be sufficient to lead towards a thermalized system of partons with signs of collectivity. This would become even more likely once softer collisions are accounted for. 
The calculations clearly indicate that parton multiplication following 
initial scattering among primary partons drives and is driven by substantially
increased multiple scatterings. The system thus created has a large
number of partons undergoing semi-hard multiple collisions and the 
multi-parton interactions rise rapidly with increase in the center of 
mass energy.

The corresponding rapidity density distributions of partons produced for the three 
sets of calculations are shown in Fig.~\ref{dndy-sqrts}. Once again we see that
semi-hard collisions among partons {\em without} radiative processes do not lead to
a substantial rise in the production of partons in $pp$ collisions. 
The radiative
processes following multiple collisions lead to an increase in parton production
by a factor of 10--20 at LHC energies. We also note that the greatly increased multiple
collisions can lead to a substantial production of strange and charm quarks as the 
centre of mass energy increases. One should note, however, that the average $p_T$ of partons included in the analysis of the upper two frames is considerably higher due to the momentum cut-off of the scattering cross section than in the lowest frame, which includes fragmented partons that can carry a significantly lower $p_T$.
The transverse momentum spectra for these calculations (for $p_T \geq$ 2 GeV) are shown in
 Fig.~\ref{spec-sqrts} and reveal a power-law behavior as expected.

The ratio of the number of strange and light quarks produced in 
such collisions, $\gamma_s$ (Eq.\ref{gamma}), is often
used as a measure of strangeness or chemical equilibration. The results for $\gamma_s$ as a function of center
of mass energy for the three sets of the calculations for semi-hard processes considered here
are shown in Fig.~\ref{gamma-sqrts}. We immediately see that fragmentations play an important
role in increasing the value of $\gamma_s$ which for the highest energy is seen to saturate 
at a value of about 0.9, suggesting that the fragmentations and enhanced 
multiple scatterings may push the system towards equilibration of strangeness even in $pp$
collisions at higher energies. 

Let us pause here to understand this large opening up of multi-partonic interactions in $pp$ 
collisions as the energy of the collision increases. 
This has several origins. First of all, we recall that the center of mass energy 
in a collision of primary partons - $\widehat{s}$ is equal to $x_1 x_2 \sqrt{s}$, where $x_i$'s stand for the
fractions of nucleon momenta carried by the partons. The lower cut-off on the transverse momentum 
for the collision requires (see e.g., Ref.~\cite{Wang:1991hta})
 that $x_1 x_2 \sqrt{s} \geq 2 p_T^\text{cut-off}$. As $0\le x_i \leq 1$, the partons 
must have $x_i \geq 2 p_T^\text{cut-off}/\sqrt{s}$ to be able to participate in the 
semi-hard partonic collisions considered here. The structure functions for gluons and sea quarks
increase with decreasing $x$ and this will bring in many more partons which can participate in the semi-hard
collisions as the center of mass energy increases.
 Secondly the parton-parton cross-sections
rise as the available center of mass energy increases.
 And lastly with the increase in the center of
 mass energy, many 
more collisions will have large momentum transfers making it possible for the fragmentation
processes to
contribute to partons multiplications. 

In Figs.~\ref{2.76_data} and \ref{7.00_data} we give a comparison of our calculations with the
prompt charm production measured by the ALICE experiment~\cite{Abelev:2012vra,ALICE:2011aa} at 2.76
and 7.00 TeV respectively. The experimental values for $D^0$ have been divided by 0.565- the fraction for fragmentation of $c$ quarks into $D^0$. We have limited the comparison to
$p_T >$  2 GeV in view of the $p_T^\text{cut-off}$ used in our calculations.  A fair agreement is
seen, though we note a definite tendency of the calculations to give a larger 
production of charm as the $p_T$ decreases, especially at the higher incident energy. This is under investigation.   


\begin{figure}
\centerline{\includegraphics*[width=8.0 cm]{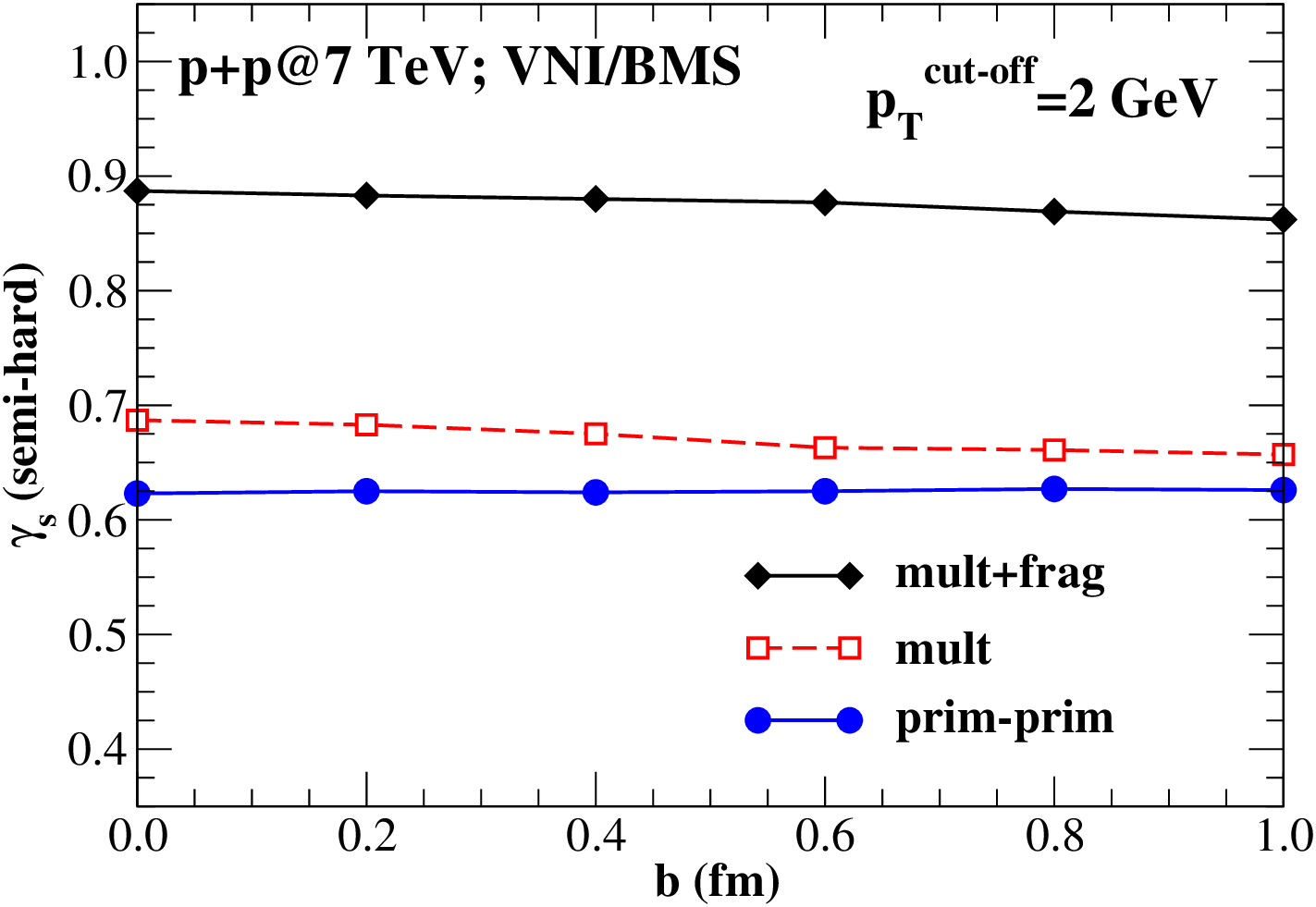}}
\caption{(Color online) The ratio of strange quarks and light quarks produced in 
semi-hard processes for the three set of calculations discussed here as a function of impact parameter.} 
\label{gamma-b}
\end{figure}

\begin{figure}
\centerline{\includegraphics*[width=8.0 cm]{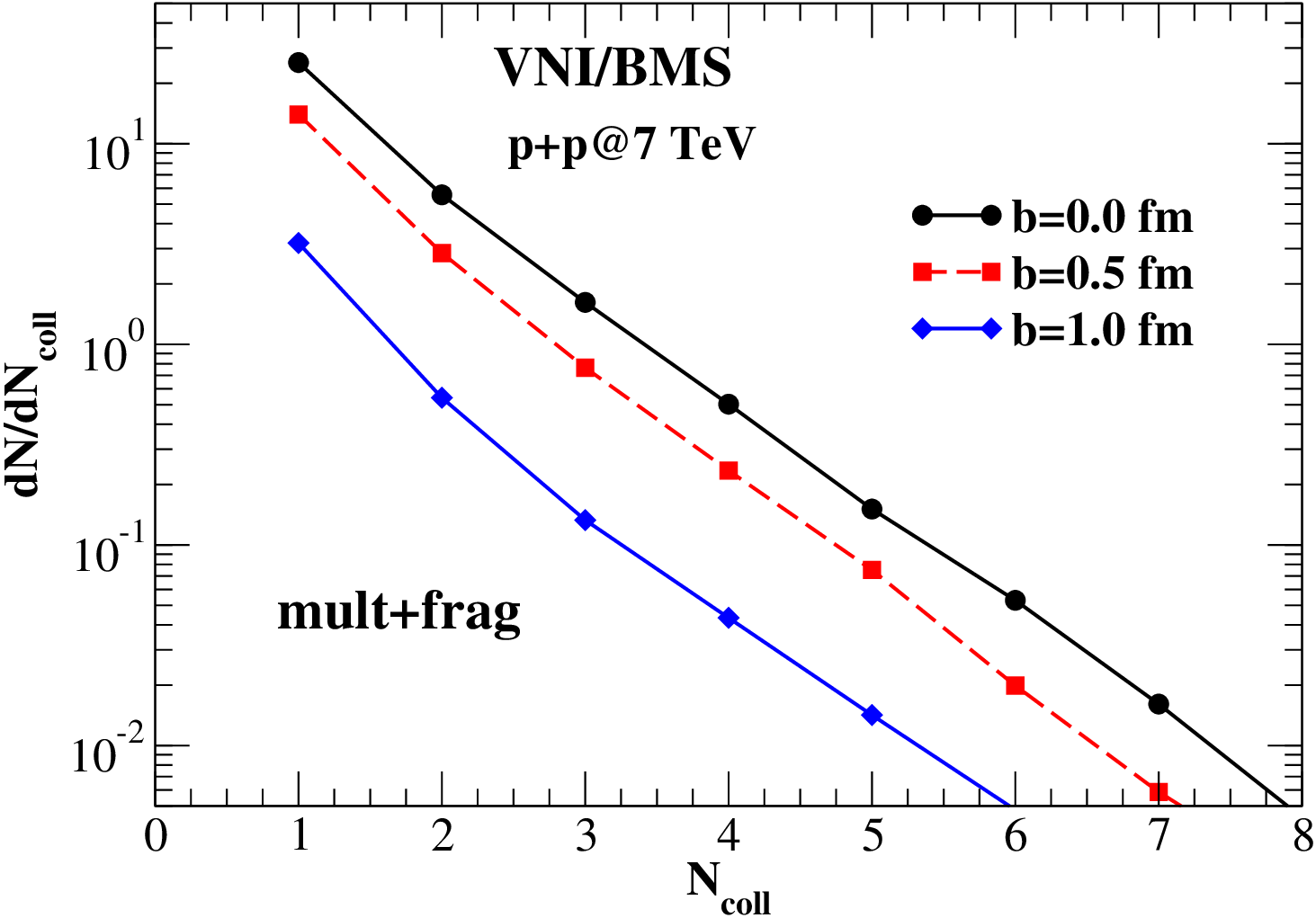}}
\centerline{\includegraphics*[width=8.0 cm]{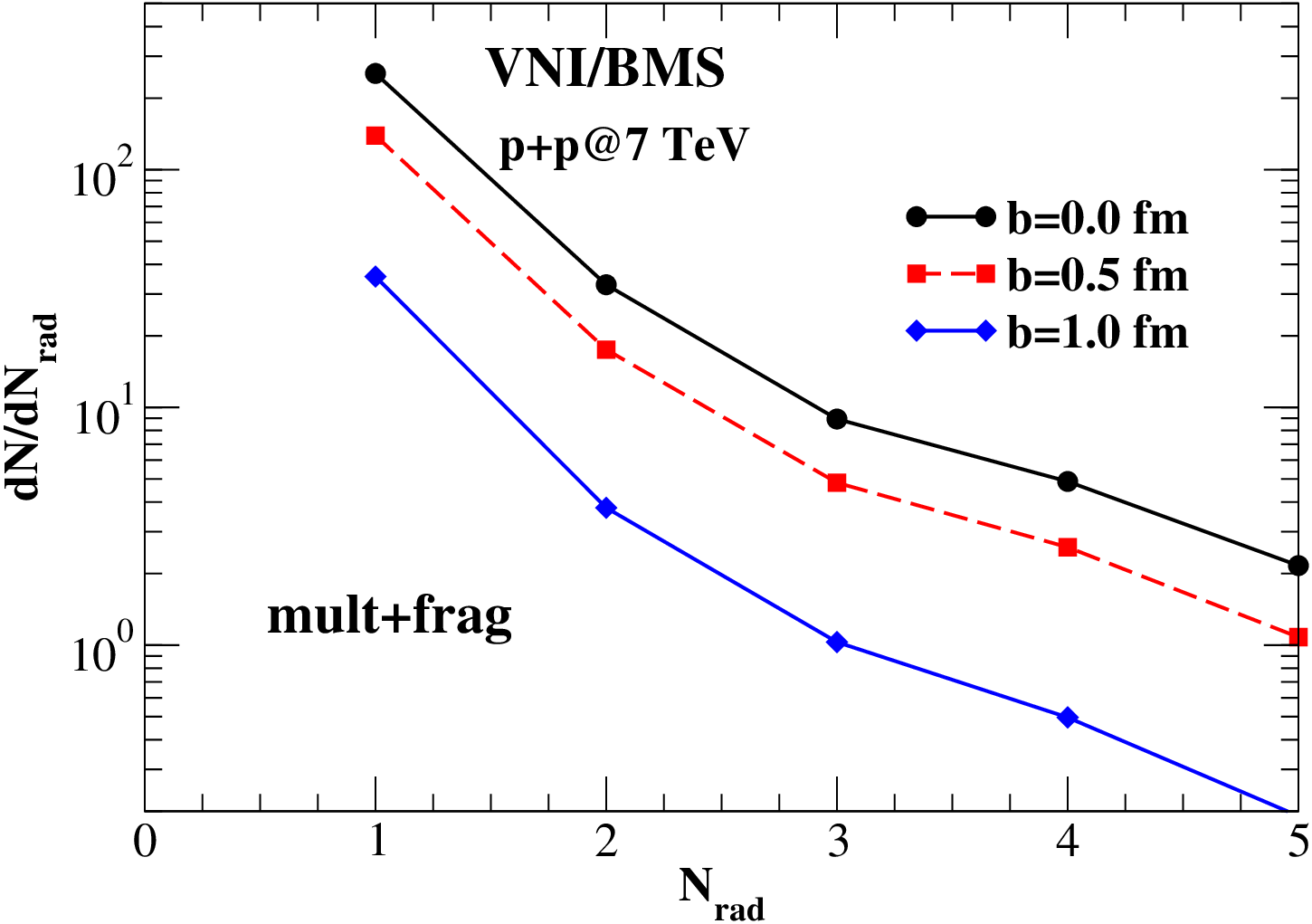}}
\caption{(Color online) The frequency distribution of number of collisions (upper panel)
and number of radiations (lower panel) suffered by partons for different impact parameters
at $\sqrt{s}$ of 7 TeV for $pp$ collisions. The multiple collisions among the
partons and radiative processes are included.} 
\label{coll_rad}
\end{figure}
\section{Evolution of partonic cascades as a function of impact parameter ($b$)  and $p_T^\text{cut-off}$}

Our calculations provide an opportunity to study the evolution of the partonic
cascade as a function of impact parameter as the number of
partons in the region of over-lap changes with the impact parameter.
 Thus for example in a collision of protons
at the center of mass  energy of 7 TeV (a case which we study in greater detail here), each proton is populated by about 270 partons,
which include the up and down valence quarks, up, down and strange sea quarks, and gluons
distributed according to the function given by Eq.~\ref{nucleon} given earlier. This
immediately provides for a larger possibility for multiple collisions for smaller impact 
parameters. 
We acknowledge that the  identification of the impact parameter in a $pp$
collision is experimentally rather challenging, yet we proceed under the assumption that some measure of centrality can be identified that allows experimental data to be mapped to our systematic study as a function of the impact parameter (over which we have full control in our calculation).

We note that our calculation contains a large number of uninteracted partons which will subsequently hadronize.
It is quite likely that the hadrons arising from these uninteracted partons will have lower transverse
momenta while those resulting from the partons which have undergone semi-hard collisions will have 
larger transverse momenta. 

We give results of our calculations for impact parameter, $b$, equal to 0.0, 0.2, 0.4, 0.6, 0.8, and 1.0 fm.
The $p_T^\text{cut-off}$ for these calculations has been fixed at 2 GeV.

Looking at the total number of light quarks, strange quarks, charm quarks, gluons and number of collisions
and fragmentations (when applicable) (see Fig~\ref{impact}), we see a very clear  increase
 in the number of collisions as the impact parameter decreases. These variations are large enough to
provide a distinctive classification of events with large semi-hard partonic collisions for the more
realistic calculation of partonic collisions along with radiative processes.

\begin{figure}
\centerline{\includegraphics*[width=8.3 cm]{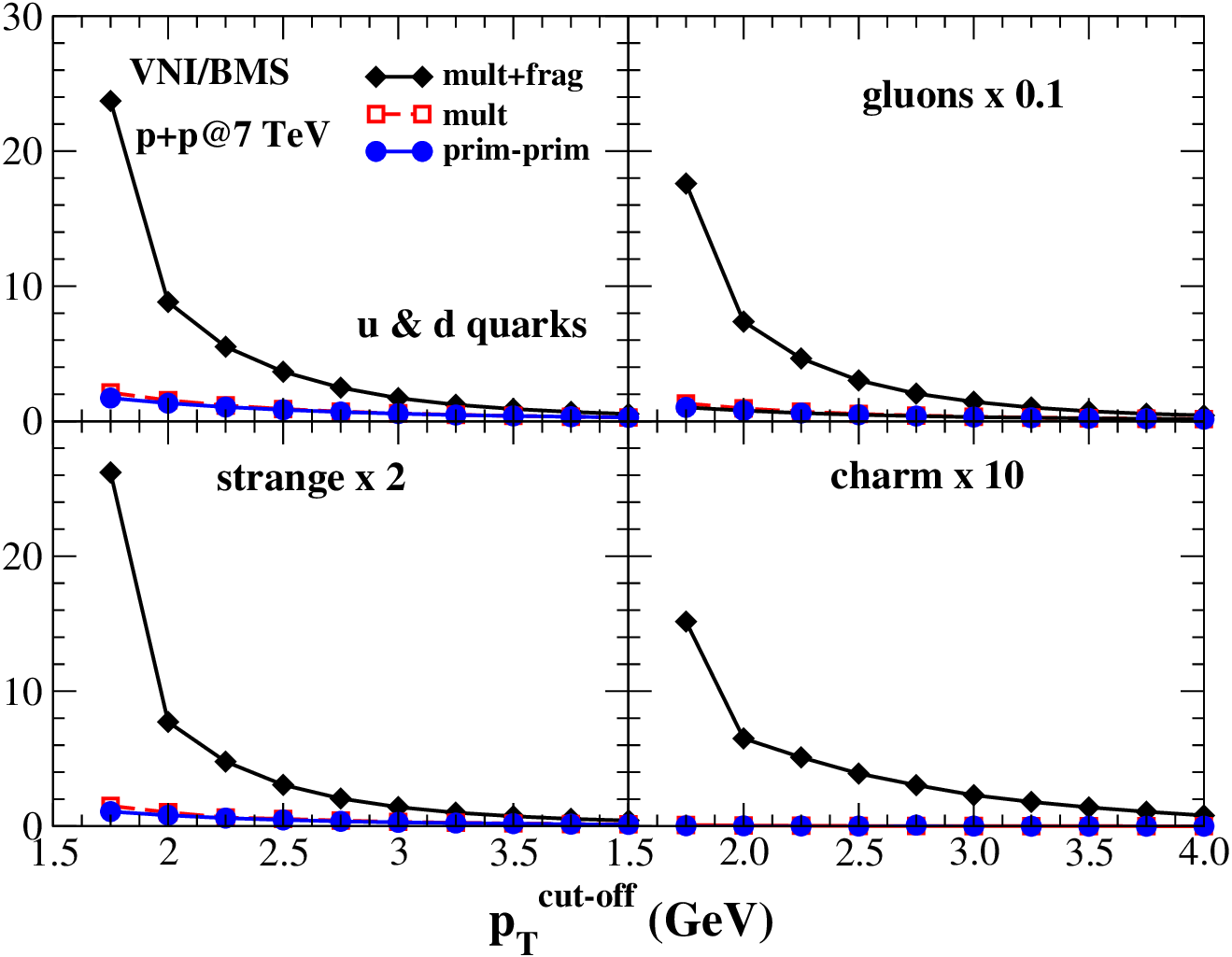}}
\centerline{\includegraphics*[width=8.3 cm]{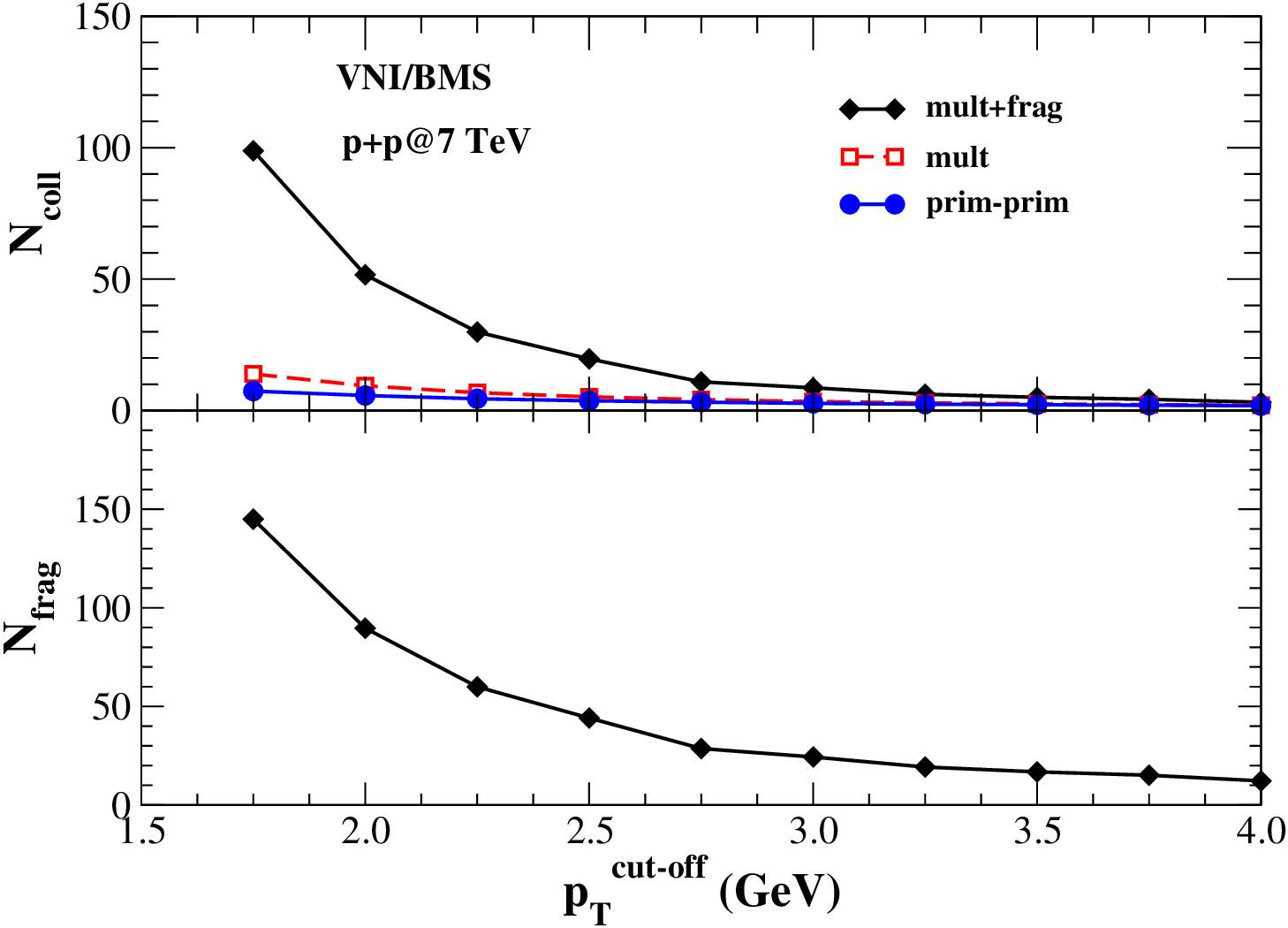}}
\caption{(Color online) [Upper panel] Effect of changing 
the lower $p_T^\text{cut-off}$ on production of light quarks, strange quarks, charm quarks and gluons 
as a function of $p_T^{\rm{cut-off}}$ for semi-hard partonic collisions in $pp$  collisions
at $\sqrt{s}=$ 7 TeV for 
calculations involving
scattering only between primary partons (filled circles), multiple scatterings {\em with-out} fragmentation
 of scattered partons (hollow squares)
and multiple scatterings {\em with} fragmentation of scattered partons (filled diamonds). [Lower panel] Number of collisions and
 number of fragmentations  as a function of $p_T^{\rm{cut-off}}$ 
set of calculations.}
\label{ptcut}
\end{figure}

How are the rapidity density and transverse momentum distributions of partons
 produced in these semi-hard processes affected by variation in impact parameter?
We give these results for the rapidity densities for the three set of calculations in Fig.\ref{dndy-b}.

\begin{figure}
\centerline{\includegraphics*[width=8.0 cm]{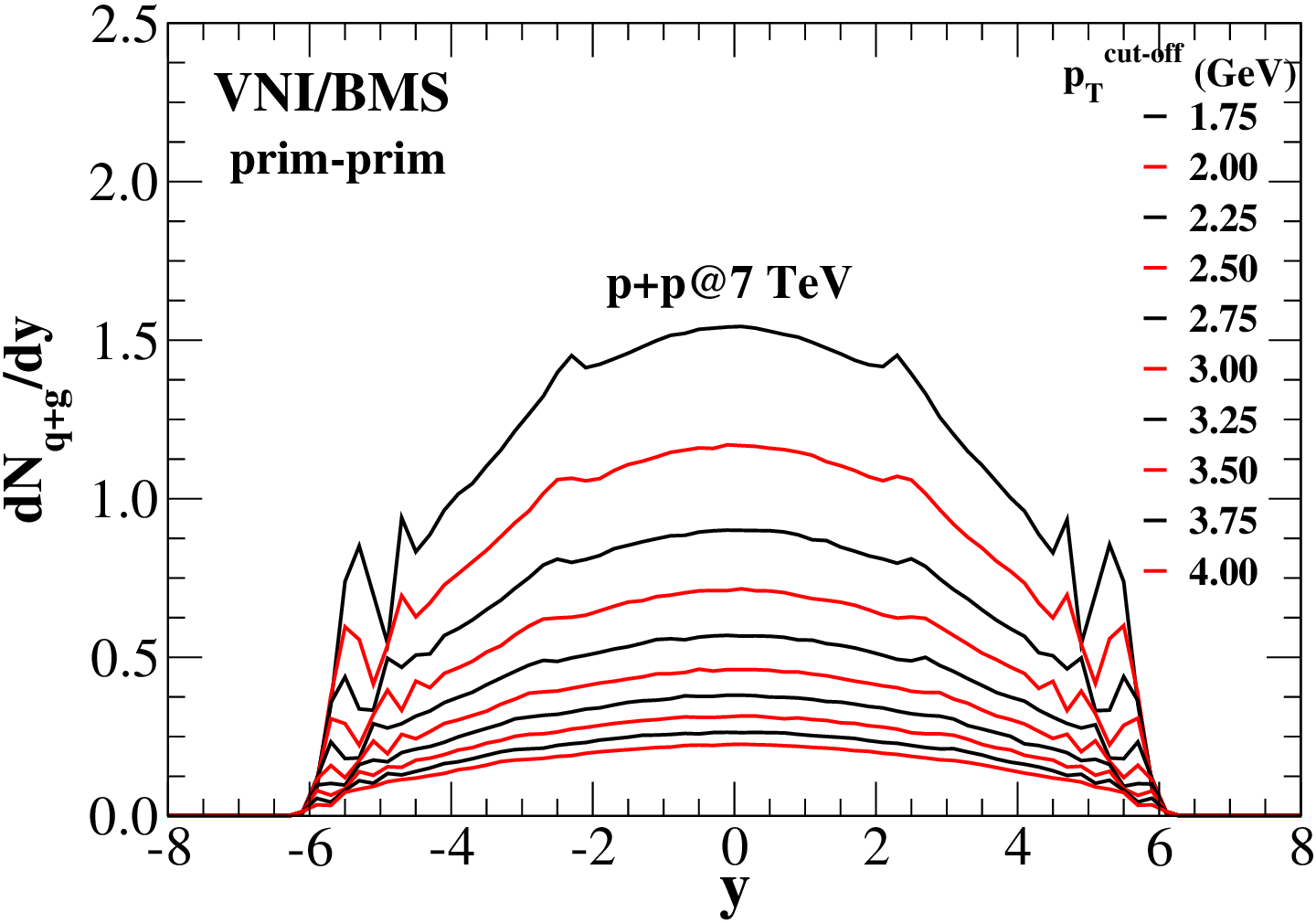}}
\centerline{\includegraphics*[width=8.0 cm]{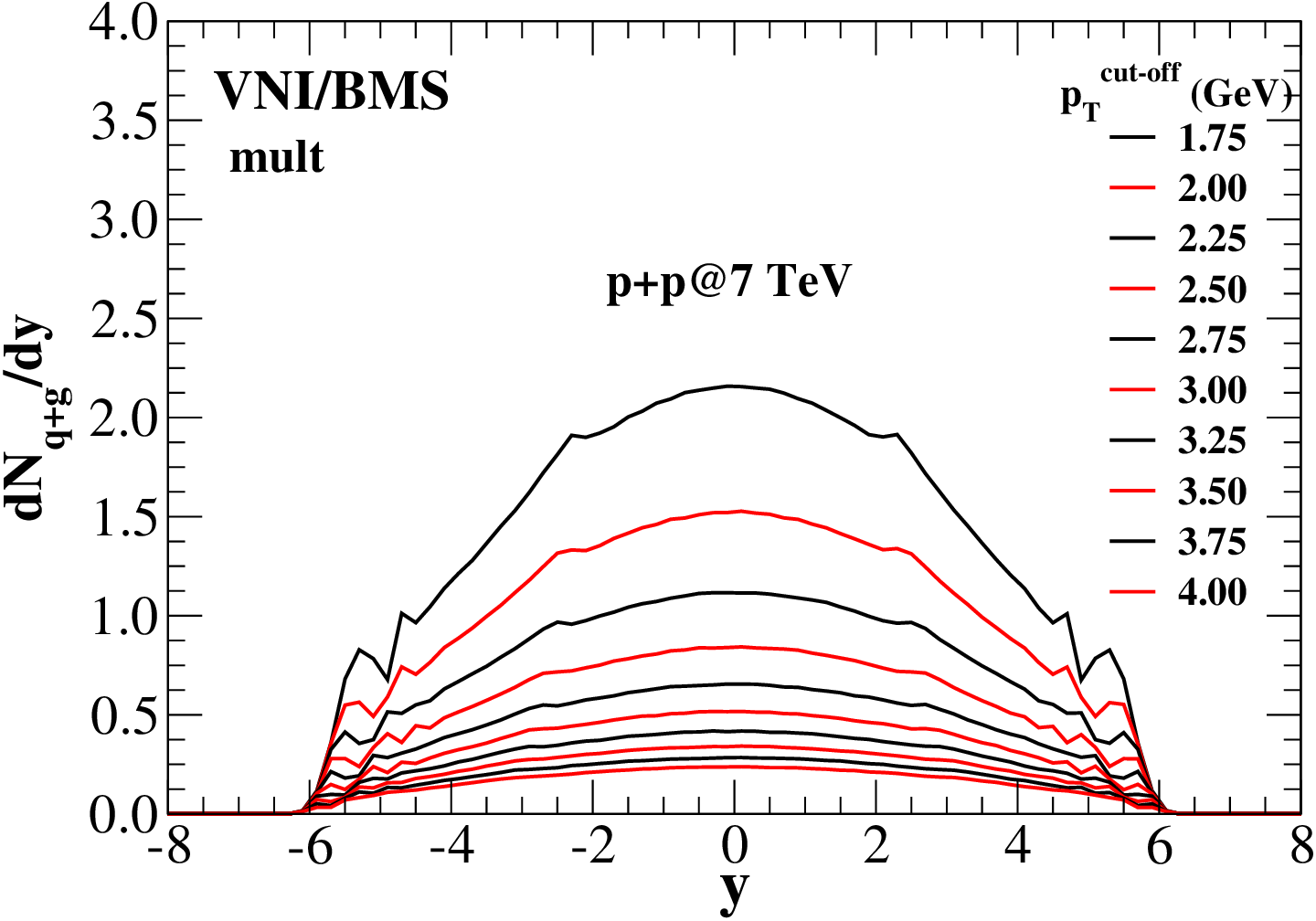}}
\centerline{\includegraphics*[width=8.0 cm]{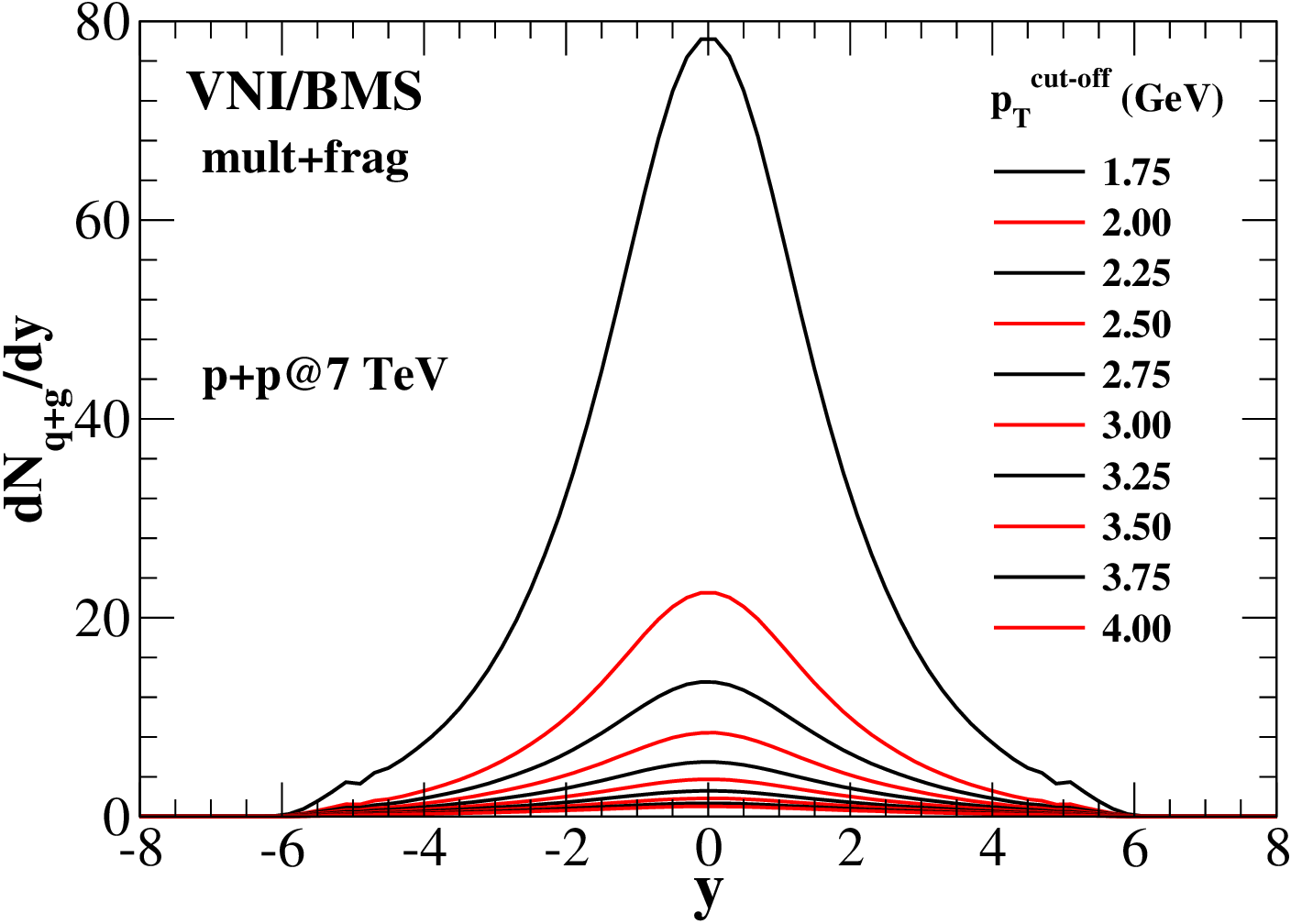}}
\caption{(Color online) Rapidity density of partons produced in $pp$ interactions at
$\sqrt{s}=$ 7 TeV
due to semi-hard collisions among primary partons (upper panel), multiple collisions
 {\em without} fragmentations of final state partons (middle
panel) and multiple collisions {\em with} fragmentation of final state
partons (lower panel) at different cut-offs for transverse momentum in semi-hard collisions.}  
\label{dndy-pt}
\end{figure}

\begin{figure}
\centerline{\includegraphics*[width=8.0 cm]{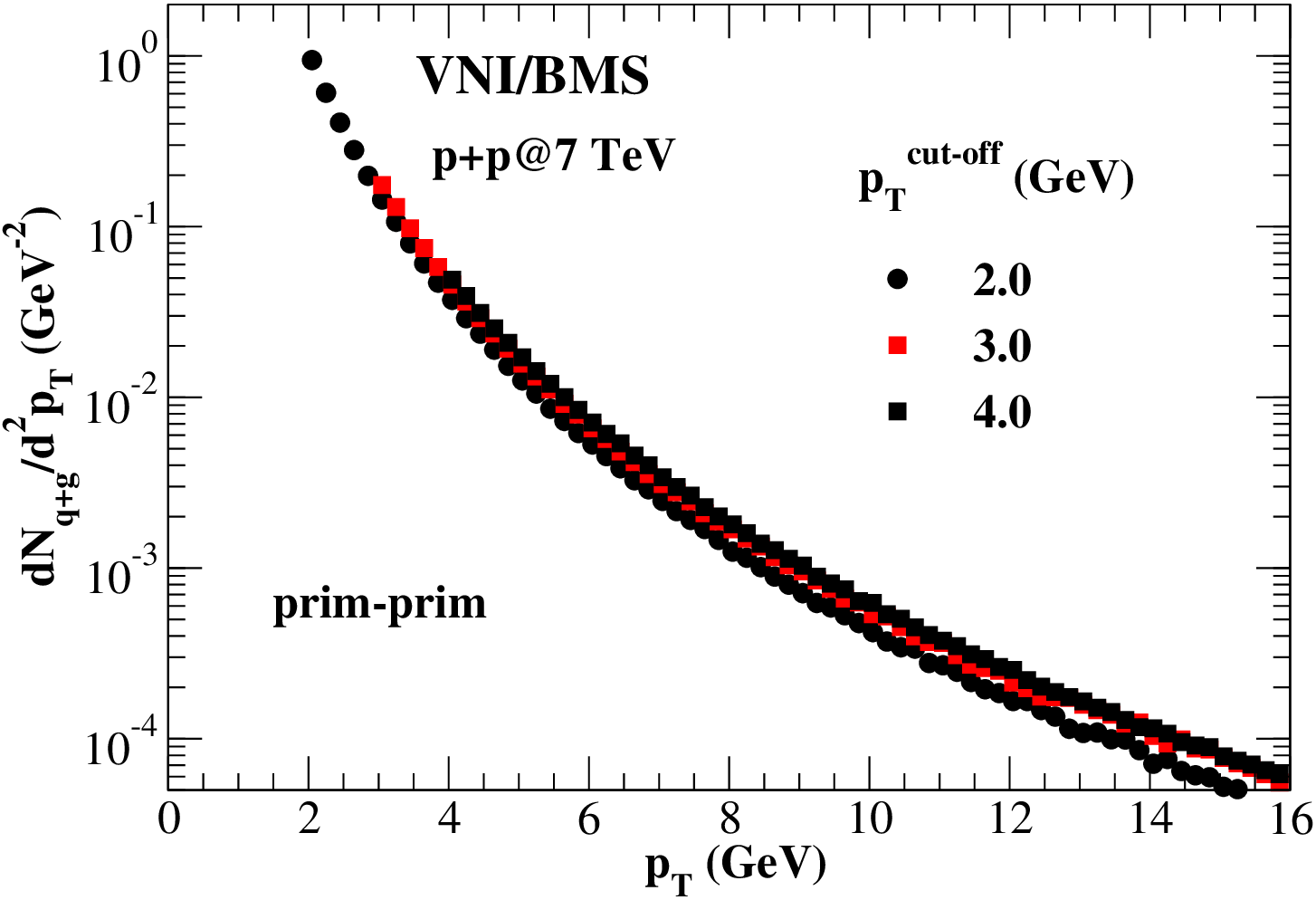}}
\centerline{\includegraphics*[width=8.0 cm]{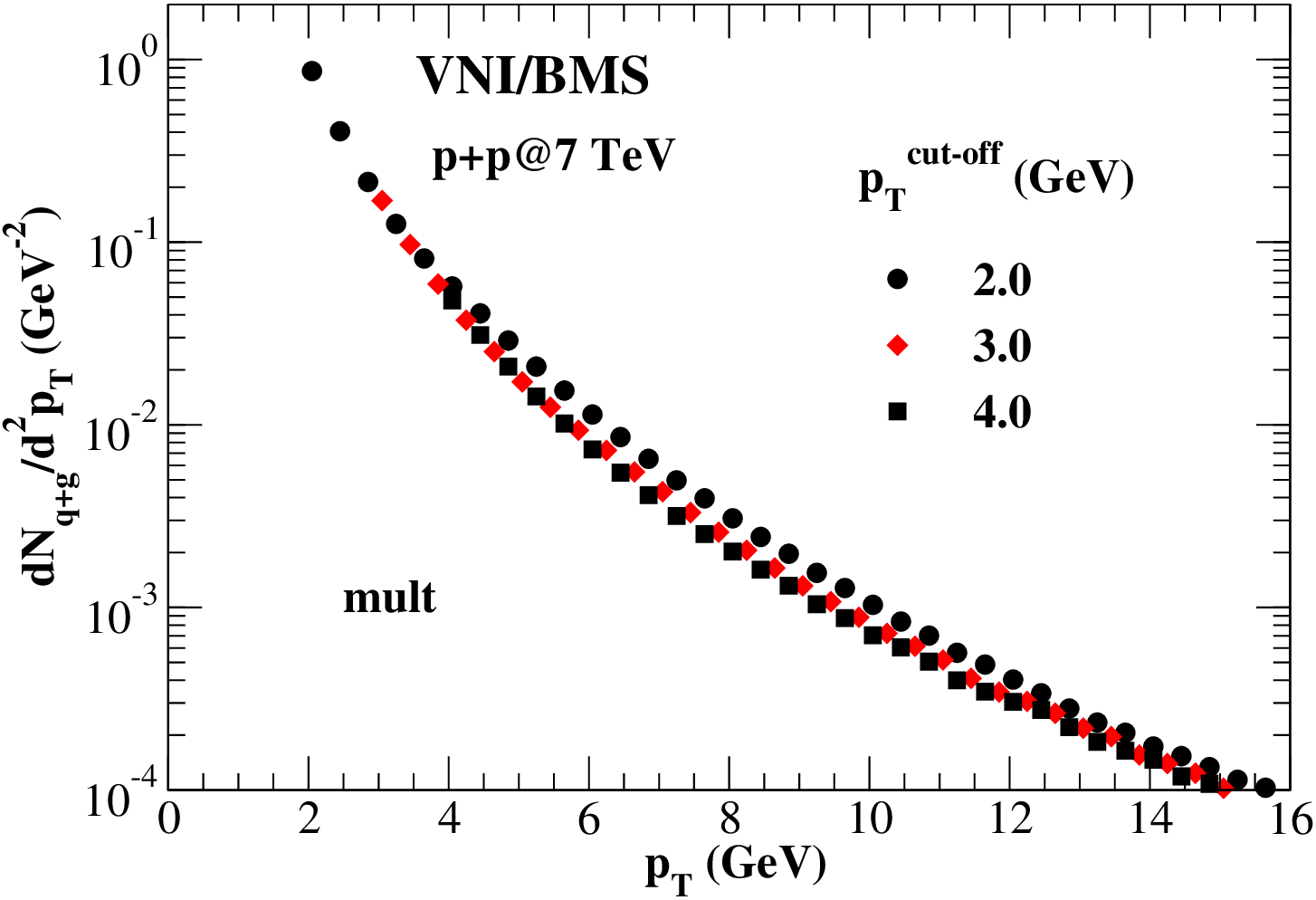}}
\centerline{\includegraphics*[width=8.0 cm]{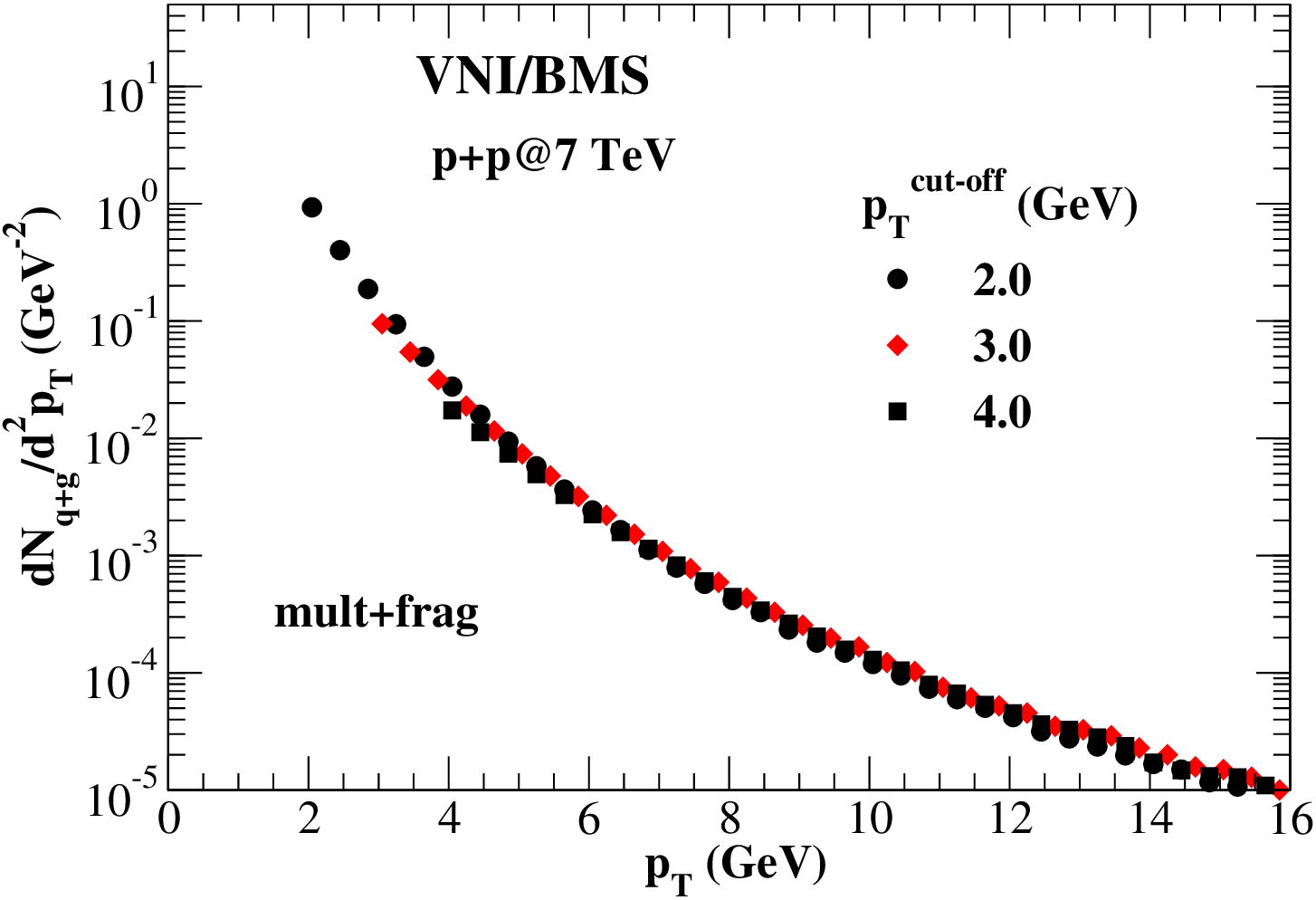}}
\caption{(Color online)Rapidity integrated $p_T$ spectra of partons produced in $pp$ interactions at
$\sqrt{s}=$ 7 TeV due to semi-hard collisions among primary partons (upper panel), multiple collisions
 {\em without} fragmentations of final state partons (middle
panel) and multiple collisions {\em with} fragmentation of final state
partons (lower panel) at different cut-offs for transverse momentum in semi-hard collisions.}  
\label{spec-pt}
\end{figure}

We see once again that multiple collisions along with parton fragmentations lead to a large production of
partons due to semi-hard processes. This production is seen to rise with decrease in impact parameter which
leads to a larger overlap of partonic clouds. The production is seen to be largest at central rapidities.
(The structures seen in the $dN/dy$ distributions for the calculations invoking only 
primary-primary or multiple collisions without fragmentations arise due to cut-offs on $p_T$.)

Fig.\ref{spec-b} shows the respective transverse momentum spectra:
we find that the distributions are quite similar for different impact parameters
but differ in magnitude. This may indicate that in high energy $pp$ interactions
even though the number of partonic collisions increases as we reduce the impact parameter,
the number of semi-hard collisions suffered by individual partons may not increase 
very substantially - as that could alter the shape of these momentum distributions. 
We shall come back to this point again.

The results for the strangeness enhancement factor $\gamma_s$ in semi-hard processes
are given in Fig.~\ref{gamma-b}. We find that $\gamma_s$ is only marginally dependent on the impact
parameter. Note, however, that even the lowest transverse momentum that we have considered is 
much larger than the mass of strange quarks.
 
The behavior of the
strangeness enhancement parameter can be understood by noting that the frequency
of number of collisions and number of radiations suffered by the parton indicate the possibility of creating
an interacting medium of partons that would allow for a measure of chemical equilibration as well.  Thus, for the most central collisions in our most realistic calculation (the scenario that includes primary and secondary scattering as well as fragmentation) we find on average in excess of 170 parton-parton scatterings and more than 300 fragmentations, even in the absence of low momentum interactions below the $p_T$-cutoff (see Fig.~\ref{coll_rad}). This figure
 shows the frequency
of number of collisions and number of radiations suffered by individual partons at three impact parameter: a substantial number of partons interact multiple times. The number of collisions as well as the parton rescattering that we observe indicate the formation of an interacting medium and a system that could potentially thermalize, as is hinted by experimental observations. Stronger statements regarding thermalization are hampered by the presence of the $p_T$-cutoff that is used to regularize the interaction cross sections in our model. 

In Fig.~\ref{ptcut}, we vary the value of $p_T^\text{cut-off}$ (for a minimum bias sample of p+p collisions) to investigate these quantities within a reasonable range of cut-off values. The observed trends (strongly rising collision and fragmentation numbers with reduced values of the cut-off) are certainly favorable for the formation of a thermalized medium.

The effect of the variation of the cut-off on 
the rapidity density distribution
 and rapidity integrated transverse momentum spectra is shown in figures~\ref{dndy-pt} and \ref{spec-pt} respectively.
 We again see a rapid rise in the parton production as the $p_T^\text{cut-off}$ 
is reduced leading to more collisions and fragmentations. We have shown only a selected set of
results for the parton spectra to avoid severe over-crowding. The results
for other values of the cut-off parameter lie between appropriate curves given here.
We see a near identity of $p_T$ 
spectra beyond the largest $p_T^\text{cut-off}$, as expected, i.e. while the low $p_T$ results of our calculation are significantly affected by the cut-off, the high momentum results remain stable.

\begin{figure}
\centerline{\includegraphics*[width=8.0 cm]{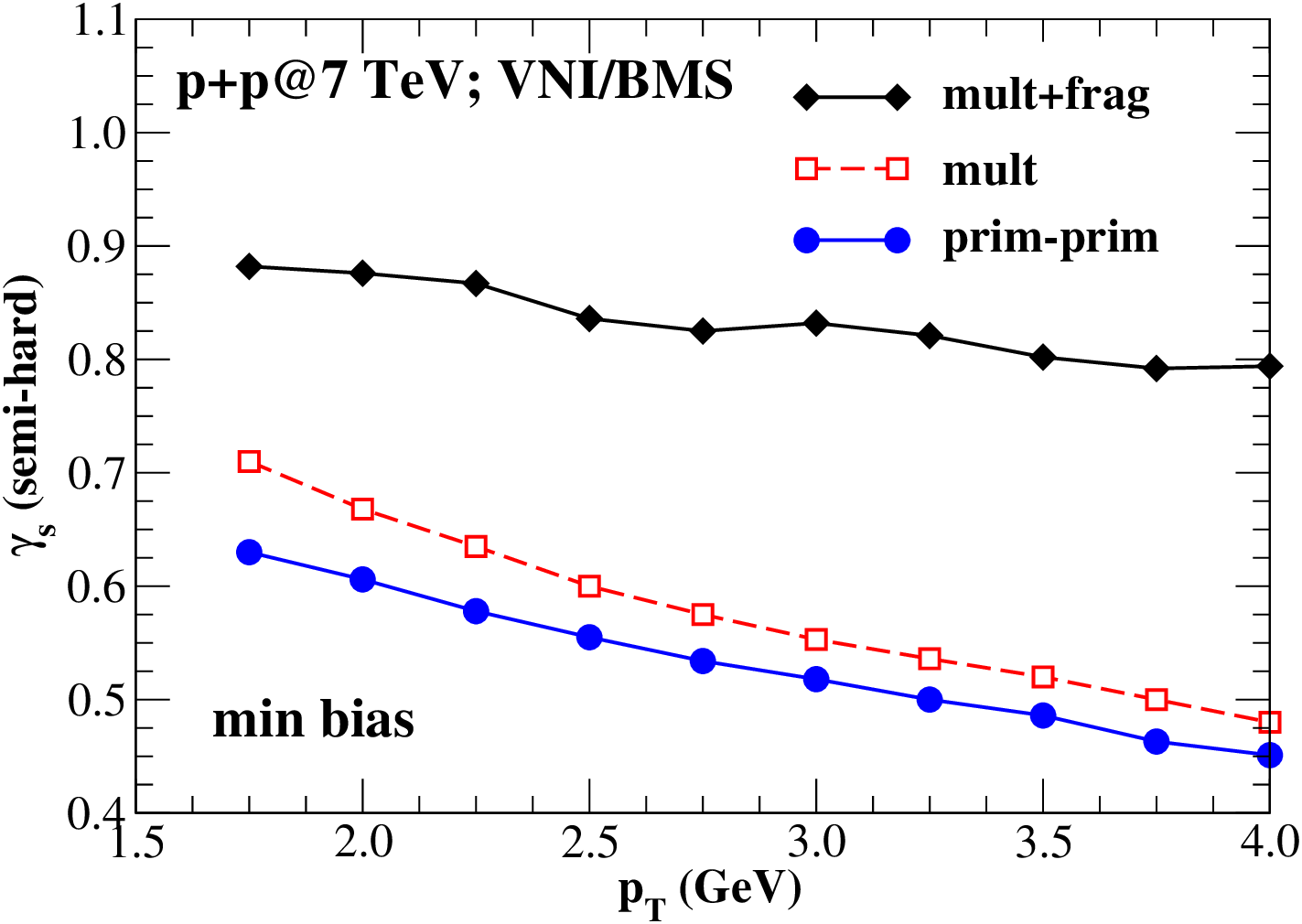}}
\caption{(Color online) The ratio of strange quarks and light quarks produced in 
semi-hard processes for the three set of calculations discussed here as a function 
of $p_T^\text{cut-off}$ in $pp$ collisions at $\sqrt{s}$ = 7 TeV.} 
\label{gamma-pt}
\end{figure}
\begin{figure}
\centerline{\includegraphics*[width=8.0 cm]{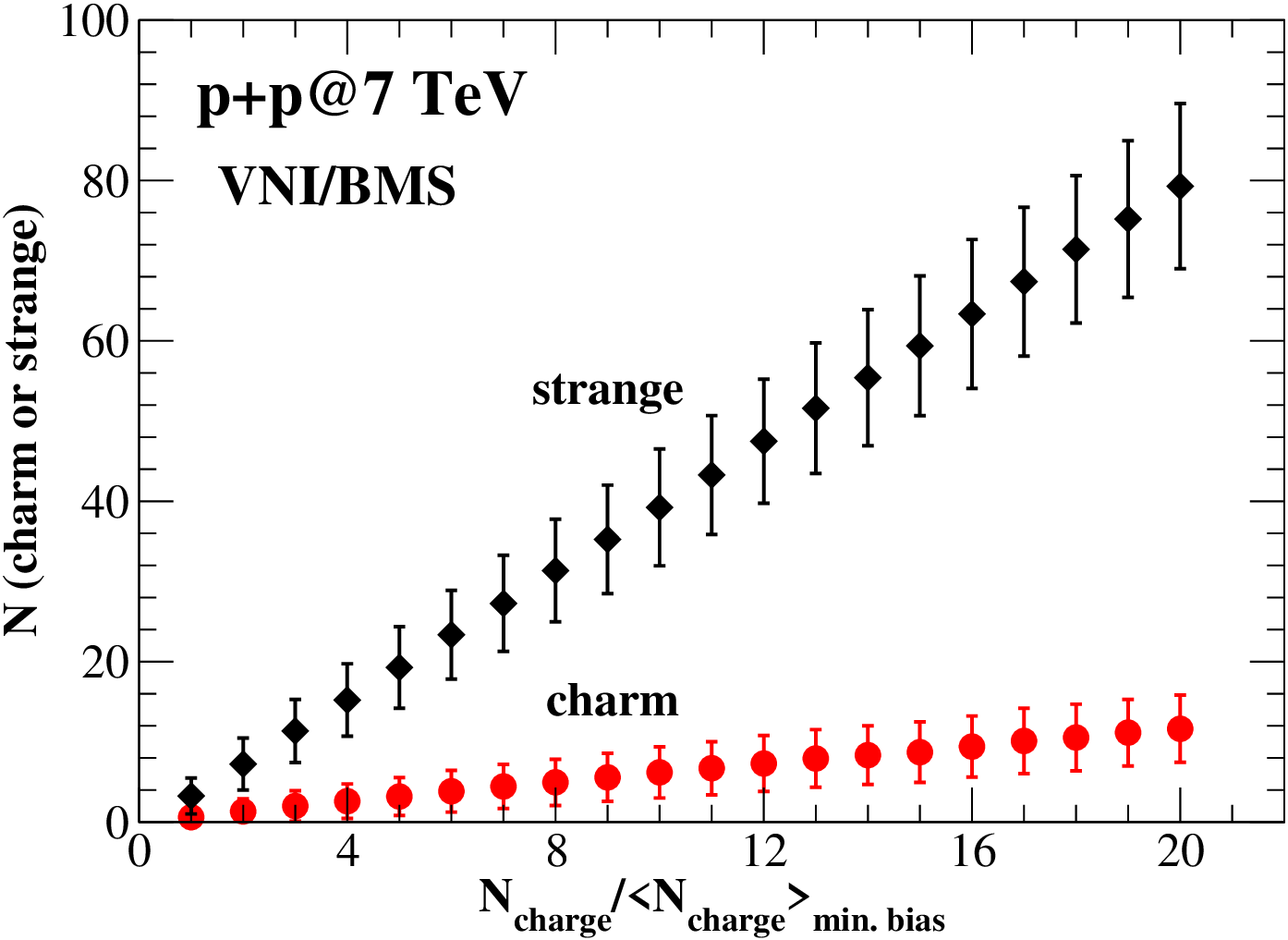}}
\centerline{\includegraphics*[width=8.0 cm]{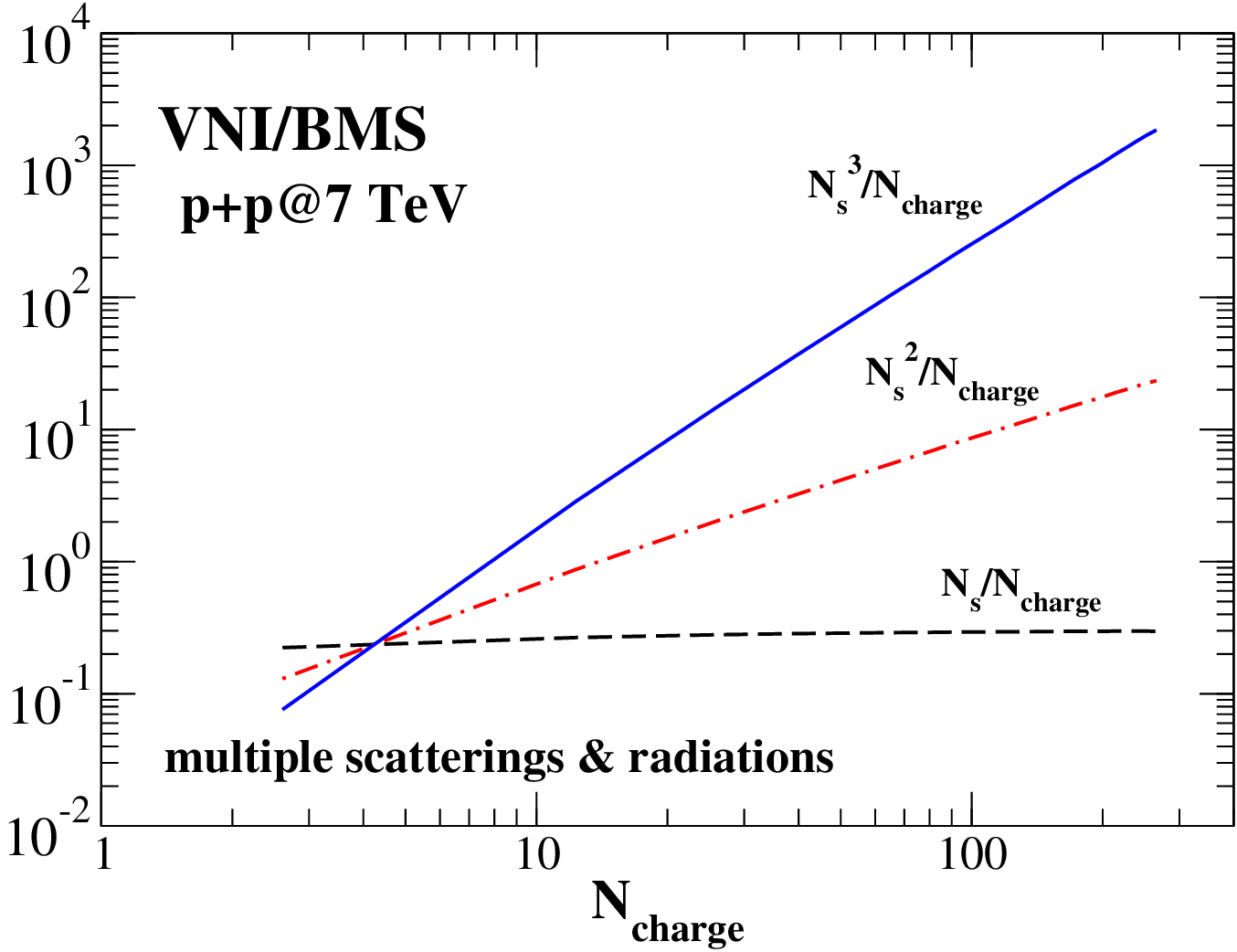}}
\caption{(Color online) Number of strange and charm quarks produced in semi-hard
interactions vs. average number of charge particles produced (upper panel).
The lower panel depicts the variation of number of strange quarks, its square and cube.}
\label{str_vs_charge}
\end{figure}

\begin{figure}
\centerline{\includegraphics*[width=8.0 cm]{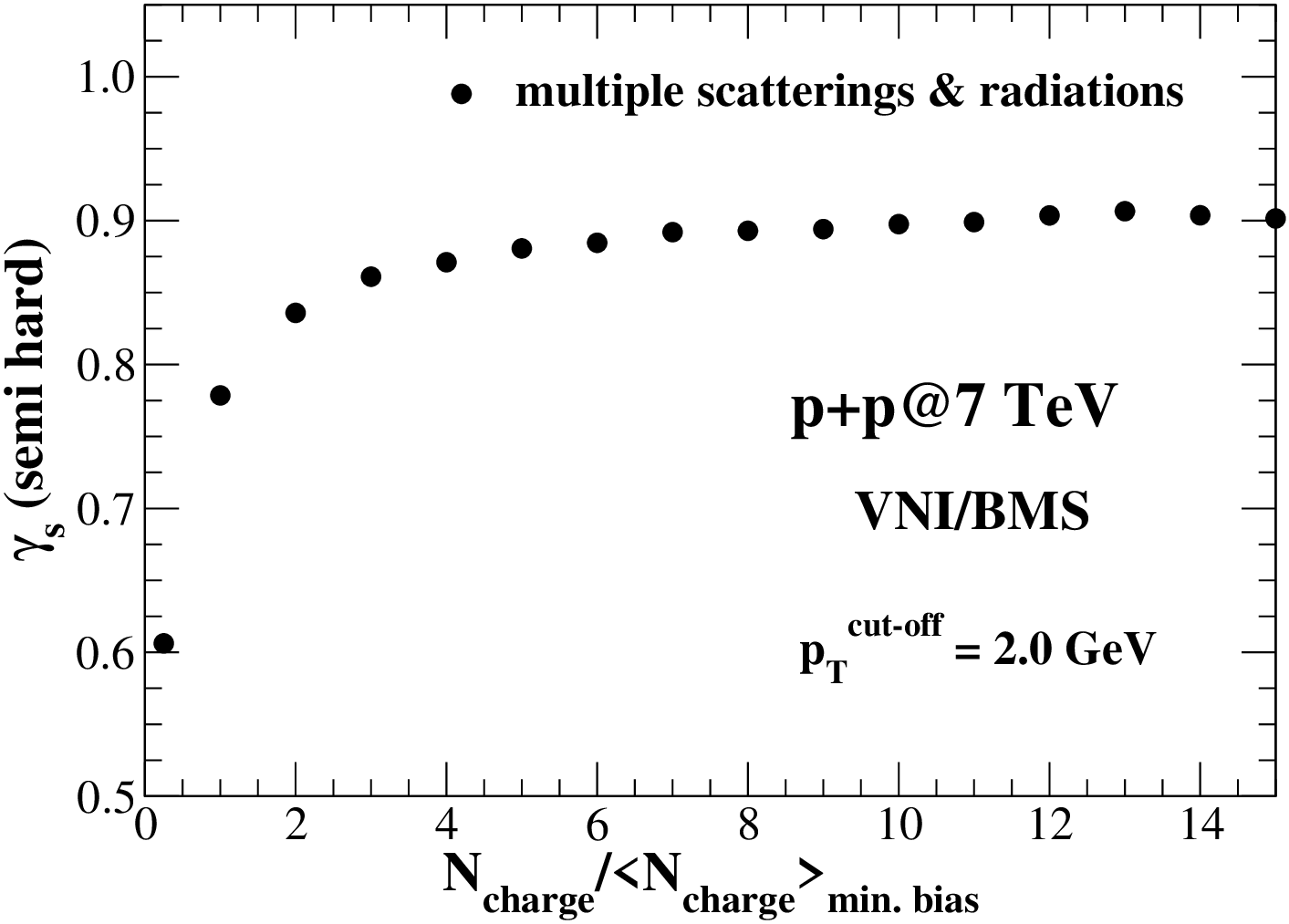}}
\caption{(Color online) The ratio of strange quarks and light quarks produced in semi-hard processes as a function of charged particles (quarks) produced. }
\label{charge_gammas}
\end{figure}

Finally the dependence of the strangeness enhancement factor $\gamma_s$ 
on the transverse momentum cut-off is
shown in Fig.~\ref{gamma-pt}. We see a mild rise in $\gamma_s$ as the 
$p_T^\text{cut-off}$ is decreased for all the calculations reported here
as these lead to increased number of partonic collisions.

\section{Evolution of strangeness for events with large multiplicity}

In the previous section we saw that production of partons rises with decrease in
impact parameter. While the determination of impact parameter for 
$pp$ collisions may be non-trivial, one can easily compare the results 
for minimum bias events to those with high multiplicity.
 
In Fig.~\ref{str_vs_charge} we show the increase in production of strangeness
and charm in collisions with large multiplicity, for the calculations
allowing multiple scattering among partons and the fragmentation of final state partons.
A rapid increase in strangeness and charm multiplicity is seen as the number of partons rises.

The lower frame of Fig.~\ref{str_vs_charge} shows the variation of the ratio of strange quarks vs. the number of quarks
(charged particles) produced against  the multiplicity of the charged particles produced
in semi-hard interactions. We find it to be essentially constant, basically as the $p_T^{\text{cut-off}}$
of 2 GeV chosen in these studies is much larger than the mass of strange quark taken in the
calculations. The behavior of the square and the cube of the number of strange quarks is also shown.
Taken together, these results indicate an increasing production of hadrons with one, two, and three
strange quarks as we look at events with larger multiplicity, as indeed observed in recent experiments
at LHC.

In Fig.~\ref{charge_gammas} we have plotted our result for strangeness enhancement factor as a function of charged particles.
 We see that as the number of quarks (charged particles) produced rises in the event, the strangeness tends to equilibrate, as indeed indicated in results from the ALICE experiment.

\section{Summary and conclusions}
We have studied the formation of a partonic medium produced in $pp$ collisions using parton cascade model.
The production of light quarks, strange quarks, charm quarks, and gluons in calculations permitting
multiple collisions and fragmentation of scattered partons is seen to rise rapidly with the energy
due to
semi-hard interactions treated using pQCD. The multi-parton interactions (collisions and radiations)
are seen to rise with decrease in impact parameter and decrease in the lower cut-off on the 
transverse momenta at a given centre of mass energy.
 The strangeness enhance factor for the semi-hard
processes at higher energies is seen to be close to unity and only marginally dependent on the impact parameter.
A detailed analysis of number of collisions and radiations suffered by partons reveals that while many
partons undergo only one collision and one fragmentation per event, the chances of the same parton 
undergoing multiple collisions or several radiations is still substantial. This indicates  the emergence of an interacting medium. The strangeness and charm production is also seen to rise as a function of charged particles
as indicated in recent experiments. The strangeness is seen to equilibrate in high multiplicity events and as the centre of mass energy increases. Overall we deem the amount of partons produced and the multiple interactions among these partons favorable for the formation of an interaction medium that may give rise to collective effects.

\section*{Acknowledgments} 
DKS gratefully acknowledges the support by the Department of Atomic
Energy. This research was supported in part by the ExtreMe Matter Institute EMMI at the GSI Helmholtzzentrum für Schwerionenforschung, Darmstadt,
Germany. DKS also acknowledges valuable comments by Helmut Satz and Dirk Rischke.
SAB acknowledges support by US Department of Energy grant DE-FG02-05ER41367.

\newpage


\end{document}